\documentclass{jfm}

\usepackage{graphicx}
\usepackage{newtxtext}
\usepackage{newtxmath}
\usepackage{natbib}
\usepackage{hyperref}
\hypersetup{
    colorlinks = true,
    urlcolor   = blue,
    citecolor  = black,
}

\newcommand{\RomanNumeralCaps}[1]
\linenumbers

\newcommand{\abs}[1]{\left \lvert #1 \right\rvert} % for absolute value
\renewcommand{\vec}[1]{\boldsymbol{#1}  } % for vectors

\renewcommand{\d}[2]{\frac{\mathrm{d} #1}{\mathrm{d} #2}} % for derivatives
\newcommand{\pd}[2]{\frac{\partial #1}{\partial #2}} % for partial derivatives
\newcommand{\grad}[1]{\vec{\nabla} #1} % for gradient
\newcommand{\lap}{\nabla^2} % for laplacian 
\newcommand{\unit}[1]{\hat{\vec{#1}}}

\usepackage[utf8]{inputenc}
\usepackage{nicefrac}

\newcommand{\keps}{k_\varepsilon}
\newcommand{\krh}{k_\mathrm{Rh}}
\newcommand{\kr}{k_r}

\title{The buoyancy staircase limit in surface quasigeostrophic turbulence}

\author{Houssam Yassin
  \corresp{\email{hyassin@princeton.edu}}}

\affiliation{Program in Atmospheric and Oceanic Sciences, Princeton University,
Princeton, NJ 08544, USA}

\begin{document}
\maketitle

\begin{abstract}
Surface buoyancy gradients over a quasigeostrophic fluid permit the existence of surface-trapped Rossby waves.
The interplay of these Rossby waves with surface quasigeostrophic turbulence results in latitudinally inhomogeneous mixing that, under certain conditions, culminates in a surface buoyancy staircase: a meridional buoyancy profile consisting of mixed-zones punctuated by sharp buoyancy gradients, with eastward jets centred at the sharp gradients and weaker westward flows in between. 
In this article, we investigate the emergence of this buoyancy staircase limit in surface quasigeostrophic turbulence and we examine the dependence of the resulting dynamics on the vertical stratification.
Over decreasing stratification [$\mathrm{d}N(z)/\mathrm{d}z\leq0$, where $N(z)$ is the buoyancy frequency], we obtain flows with a longer interaction range (than in uniform stratification) and highly dispersive Rossby waves. In the staircase limit, we find straight jets that are perturbed by eastward propagating along jet waves, similar to two-dimensional barotropic $\beta$-plane turbulence. In contrast, over increasing stratification [$\mathrm{d}N(z)/\mathrm{d}z\geq0$], we obtain flows with shorter interaction range and weakly dispersive Rossby waves. In the staircase limit, we find sinuous jets with large latitudinal meanders whose shape evolves in time due to the westward propagation of weakly dispersive along jet waves.
These along jet waves have larger amplitudes over increasing stratification than over decreasing stratification, and, as a result, the ratio of domain-averaged zonal to meridional speeds is two to three times smaller over increasing stratification than over decreasing stratification. Finally, we find that, for a given Rhines wavenumber, jets over increasing stratification are closer together than jets over decreasing stratification.
\end{abstract}

\begin{keywords}

\end{keywords}

%{\bf MSC Codes }  {\it(Optional)} Please enter your MSC Codes here

\section{Introduction}

Perturbations to a barotropic (i.e., depth-independent) fluid with a background potential vorticity gradient, $\beta>0$, propagate westward as Rossby waves.
%In the presence of a positive background potential vorticity gradient, $\beta>0$, perturbations to a barotropic (i.e., depth-independent) fluid propagate westward as Rossby waves. 
In a turbulent flow, the non-linear interplay between Rossby waves and turbulence results in the latitudinally inhomogeneous mixing of potential vorticity, which, through a positive dynamical feedback, spontaneously reorganizes the flow into one characterized by eastward jets  \citep{dritschel_multiple_2008}.
%because the Rossby wave restoring mechanism is itself dependent on the background potential vorticity, this latitudinally inhomogeneous mixing results a positive feedback under which eastward jets spontaneously emerge \citep{dritschel_multiple_2008}.
 The ultimate limit of such inhomogeneous mixing, which can be achieved for sufficiently large values of $\beta$, is a potential vorticity staircase: a piecewise constant potential vorticity profile consisting well-mixed regions separated by isolated discontinuities, with eastward jets centred at the  discontinuities and westward flows in between \citep{danilov_scaling_2004,dunkerton_barotropic_2008,scott_structure_2012,scott_zonal_2019}. 

Analogously, a buoyancy gradient at the surface of a quasigeostrophic fluid supports the existence of surface-trapped Rossby waves that are less dispersive than their barotropic counterparts \citep{held_surface_1995,lapeyre_surface_2017}. The purpose of this article is to investigate the formation of zonal jets in the presence of a background surface buoyancy gradient and to examine the realizability of surface buoyancy staircases in the surface quasigeostrophic model.
 Although the present study is the first to systematically investigate the emergence of surface quasigeostrophic jets, there are previous studies which make use of the uniformly stratified surface quasigeostrophic model with a background buoyancy gradient. 
 These include \cite{smith_turbulent_2002}, who derive the dependence of the diffusion coefficient of a passive tracer in the presence a background buoyancy gradient. 
 Another is \cite{sukhatme_local_2009}, who, in their investigation of $\alpha$-turbulence models with a background gradient, note that, because of the decreased interaction range, surface quasigeostrophic jets in uniform stratification should be narrower than their counterparts in the barotropic model. 
 Finally, \cite{lapeyre_surface_2017} demonstrates that jets can indeed form in the uniformly stratified surface quasigeostrophic model.
 
We also investigate how surface quasigeostrophic jets depend on the underlying vertical stratification. \cite{yassin_surface_2022} show that the vertical stratification modifies the interaction range of vortices in the surface quasigeostrophic model. Suppose we have an infinitely deep fluid governed by the time-evolution of geostrophic buoyancy anomalies at its upper boundary. Then if the stratification is decreasing [$N'(z)\leq0$, where $N(z)$ is buoyancy frequency] towards the fluid's surface (that is, the upper boundary), then the interaction range is longer than in the uniformly stratified model and the resulting turbulence is characterized by thin buoyancy filaments \textemdash{} analogous to the thin vorticity filaments in two-dimensional barotropic turbulence. Conversely, if the stratification is increasing [$N'(z)\geq0$] towards the surface, then the interaction range is shorter than in uniform stratification, and the buoyancy field appears spatially diffuse and lacks thin filamentary structures. In this article, we find  that the interaction range is related to Rossby wave dispersion: flows with a longer interaction range have more dispersive Rossby waves whereas flows with a shorter interaction range have less dispersive Rossby waves. One of our aims is to characterize the dependence of surface quasigeostrophic jets on the functional form of the vertical stratification.  

There are two motivations behind the present work. The first is its potential relevance to the upper ocean. Buoyancy anomalies at the ocean's surface are governed by the surface quasigeostrophic model \citep{lapeyre_dynamics_2006,lacasce_estimating_2006,isernfontanet_potential_2006}. Both numerical \citep{isernfontanet_three-dimensional_2008,lapeyre_what_2009,qiu_reconstructability_2016,qiu_reconstructing_2020,miracca-lage_can_2022} as well as observational \citep{gonzalez-haro_global_2014} studies indicate that a significant fraction of the surface geostrophic velocity is induced by sea surface buoyancy anomalies, especially over wintertime extratropical currents. Moreover, upper ocean turbulence has been found to be anisotropic \citep{maximenko_observational_2005,scott_zonal_2008}, with significant differences in anisotropy between major extratropical currents and other regions in the ocean \citep{wang_anisotropy_2019}. However, our neglect of the planetary $\beta$ effect, as well our assumption of vanishing interior potential vorticity, may limit the direct relevance of this study to the upper ocean.

%\cite{yassin_surface_2022} have shown that the interaction range of surface buoyancy anomalies is seasonal, with longer interaction ranges in wintertime when mixed-layer are deep and shorter interaction range in summer when mixed-layers are shallow. This article addresses how these buoyancy anomalies behave in the presence of a background surface buoyancy gradient, a persistent feature of the surface ocean. However, out neglect of interior potential vorticity anomalies, which are important for mixed-layer instability \citep{callies_seasonality_2015} may limit the applicability of these results to the upper ocean.
%The presence of a meridional buoyancy gradient at the ocean's surface then raises the question of whether the dynamics of these buoyancy anomalies is anisotropic, and whether this anisotropy is seasonal. However, we note our model neglects interior potential vorticity anomalies, which limits the validity of this model for the upper ocean.

The second motivation is that the variable stratification surface quasigeostrophic model is a simple two-dimensional model in which we can investigate how jet dynamics depend on the stratification's vertical structure. 
Another such model is the equivalent barotropic model for which the deformation radius represents the rigidity of the free surface. Small values of the deformation radius lead to a pliable free surface allowing a significant degree of horizontal divergence. 
The resulting flow then has an exponentially short interaction range, with a horizontal attenuation on the order of the deformation radius \citep{polvani_two-layer_1989}, and with approximately non-dispersive Rossby waves. Consequently, for a finite deformation radius, we obtain jets whose width is on the order of the deformation radius with a fixed meandering shape \citep{scott_spacing_2022}. 
In contrast, for the variable stratification surface quasigeostrophic model, rather than just specifying a constant (i.e., the deformation wavenumber), one instead has to specify the stratification's functional form, $N(z)$. 
Over decreasing stratification $[N'(z) <0]$, because of the longer interaction range and the more dispersive waves, we obtain jets similar to the two-dimensional barotropic model. 
Conversely, over increasing stratification $[N'(z)>0]$, the shorter interaction range along with the weakly dispersive waves lead to sinuous jets whose shape evolves in time through the propagation of weakly dispersive along jet waves. 
Moreover, because of these along jet waves, a smaller fraction of the total energy is contained in the zonal mode over increasing stratification (with a shorter interaction range) than over decreasing stratification (with a longer interaction range). 

The remainder of this article is organized as follows.
 Section \ref{S-theory} introduces the variable stratification surface quasigeostrophic model and shows how the stratification's vertical structure controls both the interaction range of point vortices as well as the dispersion of surface-trapped Rossby waves. 
 Then, in section \ref{S-staircase_theory}, we introduce two wavenumbers, $\keps$ and $k_r$, whose ratio, $\keps/k_r$, forms the key non-dimensional parameter of this study; here, $\keps$ is a wavenumber depending on the energy injection rate whereas $k_r$ is a wavenumber depending on surface damping rate. This non-dimensional number is a generalization of the non-dimensional number used in previous studies \citep{danilov_rhines_2002,sukoriansky_arrest_2007,scott_structure_2012}. 
 By considering an idealized buoyancy staircase, we also investigate  how the Rhines wavenumber relates to the jet spacing under decreasing, increasing, and uniform stratification.
 %We also investigate how the Rhines wavenumber relates to the jet spacing by considering an idealized buoyancy staircase. 
 Section \ref{S-numerical} then presents numerical experiments detailing the emergence of the staircase limit as $\keps/k_r$ is increased for various stratification profiles. 
 In addition, we also present experiments where we fix the external parameters and vary the vertical stratification alone.
 Finally, we conclude in section \ref{S-conclusion}.

\section{The interaction range and wave dispersion}\label{S-theory}

\subsection{Equations of motion}
 Consider an infinitely deep fluid with zero interior potential vorticity. The geostrophic streamfunction, $\psi$, then satisfies
 \begin{equation}\label{eq:zero_pv}
  \pd{}{z}\left(\frac{1}{\sigma^2} \pd{\psi}{z} \right) + \lap \psi = 0
 \end{equation}
 in the fluid interior, $z \in (-\infty,0)$. The horizontal Laplacian is denoted by $\lap = \p_x^2 + \p_y^2$ and the non-dimensional stratification is given by 
 \begin{equation}
 	 \sigma(z) = N(z)/f,
 \end{equation}
 where $N(z)$ is the buoyancy frequency and $f$ is the constant local value of the Coriolis parameter.
  Time-evolution is  determined by the material conservation of surface potential vorticity \citep{bretherton_critical_1966},
   \begin{equation}\label{eq:theta_inversion}
 	\theta = -\frac{1}{\sigma_0^2} \pd{\psi}{z}\Big|_{z=0},
  \end{equation}
  at the upper boundary, $z=0$, where $\sigma_0 = \sigma(0)$. Explicitly, the time-evolution equation is
  %% Explicitly write out D as linear damping and small scale dissipation
 \begin{equation}\label{eq:time-evolution}
 	\pd{\theta}{t} + J(\psi,\theta) + \Lambda \, \partial_x \theta = F-D,
 \end{equation}
 at $z=0$, where $J(\psi,\theta) = \partial_x \psi \, \partial_y \theta - \partial_x \theta \, \partial_y \, \psi$ represents the advection of $\theta$ by the geostrophic velocity, $\vec u = \unit z \times \grad \psi$. The frequency, $\Lambda$, is given by
 \begin{equation}
 	\Lambda = \frac{1}{\sigma_0^2} \d{U}{z}\Big|_{z=0},
 \end{equation}
 where $U(z)$ is a background zonal geostrophic flow. Without loss of generality, we have assumed that $U(0)=0$ in the time-evolution equation \eqref{eq:time-evolution} to eliminate a constant advective term. The dissipation, $D$, consists of linear damping and small-scale dissipation,
 \begin{equation}
 	D = r \, \theta + \mathrm{ssd},
 \end{equation}
 where $r$ is the damping rate.
 The forcing, $F$, and the small-scale dissipation, $\mathrm{ssd}$, are described in section \ref{S-numerical}.
  %Finally, we assume all solutions satisfy $\psi \rightarrow 0$ as $z \rightarrow -\infty$.
 
  The surface buoyancy anomaly, $b|_{z=0}$, is related to the surface potential vorticity, $\theta$, through 
  \begin{equation}
  	 b|_{z=0} = - f\,\sigma_0^2 \, \theta.
  \end{equation}
  Therefore, the time-evolution equation \eqref{eq:time-evolution} equivalently states that surface buoyancy anomalies are materially conserved in the absence of forcing and dissipation. In addition, the frequency, $\Lambda$, corresponds to a meridional buoyancy gradient,
 \begin{equation}\label{eq:dBdy}
 	\d{B}{y}\Big|_{z=0} = - f \, \sigma_0^2 \, \Lambda,
 \end{equation}
 where $B(y,z) $ is the buoyancy field that is in geostrophic balance with background zonal velocity, $U(z)$.
 
 If we further assume a doubly periodic domain in the horizontal, then we can expand the streamfunction as
 \begin{equation}\label{eq:fourier_psi}
 	\psi(\vec r, z,t) = \sum_{\vec k} \hat \psi_{\vec k}(t) \, \Psi_k(z)\, \mathrm{e}^{\mathrm{i} \vec k \cdot \vec{x}},
 \end{equation}
 where $\vec x=(x,y)$ is the horizontal position vector, $z$ is the vertical coordinate, $\vec k = (k_x,k_y)$ is the horizontal wavevector, $k=\abs{\vec k}$ is the horizontal wavenumber, and $t$ is the time coordinate. The non-dimensional wavenumber-dependent vertical structure, $\Psi_k(z)$, is determined by the boundary value problem \citep{yassin_surface_2022}
 \begin{equation}\label{eq:vertical_structure}
 	-\d{}{z} \left(\frac{1}{\sigma^2}\d{\Psi_k}{z}\right) + k^2 \, \Psi_k(z) = 0,
 \end{equation}
with the upper boundary condition
\begin{equation}\label{eq:upper}
	\Psi_k(0)=1,
\end{equation}
 and lower boundary condition
 \begin{equation}\label{eq:lower}
 	\Psi_k \rightarrow 0 \quad \textrm{as } \quad z \rightarrow -\infty.
 \end{equation}
 The upper boundary condition \eqref{eq:upper} is a normalization for the vertical structure, $\Psi_k(z)$, chosen so that 
 \begin{equation}
 	 \psi(\vec r, z=0,t) = \sum_{\vec k} \hat \psi_{\vec k}(t) \, \mathrm{e}^{\mathrm{i} \vec k \cdot \vec{x}}.
 \end{equation}
  The corresponding Fourier expansion of the surface potential vorticity is given by
  \begin{equation}
  	\theta(\vec r, t) = \sum_{\vec k} \hat \theta_{\vec k}(t) \, \mathrm{e}^{\mathrm{i} \vec k \cdot \vec{x}},
  \end{equation}
  where 
  \begin{equation}\label{eq:inversion}
  	\hat \theta_{\vec k} = - m(k) \, \hat \psi_{\vec k},
  \end{equation}
  and the function $m(k)$ is given by
  \begin{equation}\label{eq:mk}
  	m(k) = \frac{1}{\sigma_0^2} \d{\Psi_k(0)}{z}.
  \end{equation}
  The function $m(k)$ relates $\hat \theta_{\vec k}$ to $\hat \psi_{\vec k}$ in the Fourier space inversion relation \eqref{eq:inversion} and so we call $m(k)$ the \emph{inversion function}.
  
	To recover the well-known case of the uniformly stratified quasigeostrophic model \citep{held_surface_1995}, set $\sigma(z) = \sigma_0$. Then the vertical structure equation \eqref{eq:vertical_structure} along with boundary conditions \eqref{eq:upper} and \eqref{eq:lower} yield the exponentially decaying vertical structure $\Psi_k(z) = \exp\left({\sigma_0\,k\,z}\right)$. On substituting $\Psi_k(z)$ into equation \eqref{eq:mk}, we obtain a linear inversion function 
	\begin{equation}\label{eq:mk_linear}
			m(k) = \frac{k}{\sigma_0}
	\end{equation}
	 and hence [from the inversion relation \eqref{eq:inversion}] a linear-in-wavenumber inversion relation $\hat \theta_{\vec k} = - (k/\sigma_0) \, \hat \psi_{\vec k}$.
	
	\subsection{The inversion function and spatial locality}
	
	\begin{figure}
  \centerline{\includegraphics[width=1.\columnwidth]{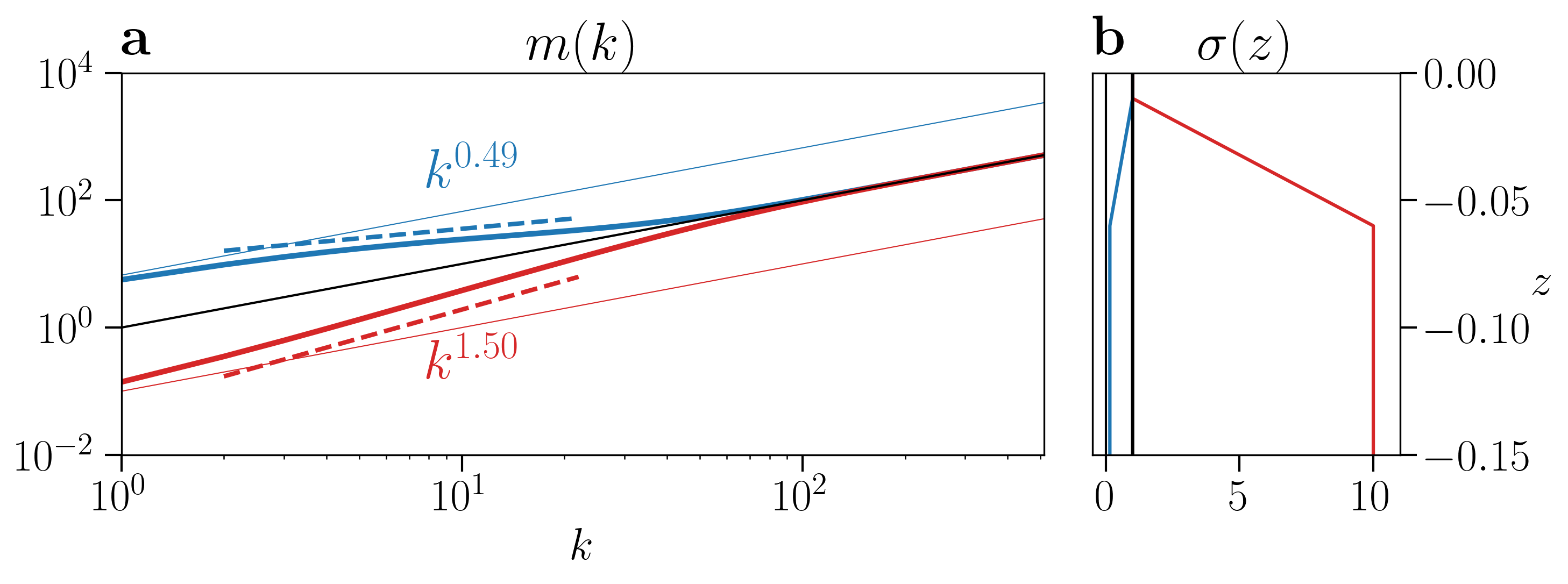}}
  \caption{The inversion functions, $m(k)$ [in panel (a)] for two stratification profiles [panel (b)] given by the piecewise stratification profile \eqref{eq:sigma_piece}. One stratification profile is increasing [$\sigma'(z)\geq 0$, blue], with $\sigma_0=1$, $\sigma_\mathrm{pyc}=0.15$, $h_\mathrm{mix} = 0.01$, and $h_\mathrm{lin}=0.05$. The other stratification profile is decreasing [$\sigma'(z)\leq 0$, red] with $\sigma_0=1$, $\sigma_\mathrm{pyc}=10$, $h_\mathrm{mix} = 0.01$, and $h_\mathrm{lin}=0.05$. The thin black line is given by $k/\sigma_0$ where $\sigma_0=1$, whereas the blue and red lines are given by $k/\sigma_\mathrm{pyc}$ with $\sigma_\mathrm{pyc}=0.15$ for the thin blue line and $\sigma_\mathrm{pyc}=10$ for the thin red line.}
  \label{F-mk_sigma}
	\end{figure}
	
	The inversion function $m(k)$, which is determined by the stratification's vertical structure, controls the spatial locality of the resulting turbulence. We illustrate this point with the following piecewise stratification profile,
	\begin{equation}\label{eq:sigma_piece}
		\sigma(z) = 
		\begin{cases}
			\sigma_0  \quad &\text{for} \quad -h_\mathrm{mix} < z < 0\\
			\sigma_0 + \Delta \sigma \left(\frac{z+h_\mathrm{mix}}{h_\mathrm{lin}}\right) \quad &\text{for} \quad -(h_\mathrm{mix} + h_\mathrm{lin}) < z < -h_\mathrm{mix}\\
			\sigma_\mathrm{pyc} \quad &\text{for}  \quad -\infty < z < -(h_\mathrm{mix} + h_\mathrm{lin}) ,
		\end{cases}	
	\end{equation}
	where $\Delta \sigma =\sigma_0 - \sigma_\mathrm{pyc}$. 
	%In the limit $h_\mathrm{lin}/h_\mathrm{mix} \ll 1$, the inversion functions takes the form \citep{yassin_surface_2022}
	%\begin{equation}\label{eq:mk_sol}
	%	m(k) = \frac{k}{\sigma_0} \left[ \frac{\cosh\left(\sigma_0h_\mathrm{mix} k\right) + \left(\frac{\sigma_\mathrm{pyc}}{\sigma_0} \right)\sinh\left(\sigma_0h_\mathrm{mix}  k\right)}{\sinh\left(\sigma_0h_\mathrm{mix}  k\right) + \left(\frac{\sigma_\mathrm{pyc}}{\sigma_0} \right)\cosh\left(\sigma_0 h_\mathrm{mix} k\right)} \right]
	%\end{equation}
	At small horizontal scales, where $k\gg k_s$, and 
	\begin{equation}
		k_s=1/\left(\sigma_0\,h_\mathrm{mix}\right),
	\end{equation}
	then $m(k) \approx k/\sigma_0$, as in the uniformly stratified model of \cite{held_surface_1995}. Likewise, in the large-scale limit, where $k \ll k_\mathrm{pyc}$, and 
	\begin{equation}
		k_\mathrm{pyc} = 
		\begin{cases}
			1/\left(\sigma_\mathrm{pyc} \, h_\mathrm{mix}\right) \quad &\text{for} \quad \Delta \sigma \leq 0 \\
			\sigma_\mathrm{pyc}/\left(\sigma_0^2 \, h_\mathrm{mix}\right) \quad &\text{for} \quad \Delta \sigma>0,
		\end{cases}
	\end{equation}
	then $m(k) \approx k/\sigma_\mathrm{pyc}$. However, for wavenumbers between $k_\mathrm{pyc} \lesssim k \lesssim k_s$, the inversion function takes an approximate power law form
	\begin{equation}\label{eq:mk_power}
		m(k) \approx m_0 \, k^{\alpha},
	\end{equation}
	where $m_0>0$ and $\alpha \geq 0$.
	The power $\alpha$ depends on the ratio $\sigma_\mathrm{pyc}/\sigma_0$ between the deep and surface stratification. If the stratification decreases towards the surface [$\sigma^\prime(z) \leq 0$, or $\sigma_\mathrm{pyc}/\sigma_0>1$] then $\alpha > 1$, with $\sigma_\mathrm{pyc}/\sigma_0 \rightarrow \infty$ sending $\alpha \rightarrow 2$. In contrast, if the stratification increases towards the surface [$\sigma^\prime(z) \geq 0$, or $\sigma_\mathrm{pyc}/\sigma_0 < 1$] then $\alpha <1$, with  $\sigma_\mathrm{pyc}/\sigma_0 \rightarrow 0$ sending $\alpha \rightarrow 0$.	
	Thus, for wavenumbers $k_\mathrm{pyc} \lesssim k \lesssim k_s$, the inversion relation \eqref{eq:inversion} has the approximate form
	\begin{equation}
		\hat \xi_{\vec k} = - k^\alpha \, \hat \psi_{\vec k},
	\end{equation}
	where $\hat \xi_{\vec{k}}= \hat \theta_{\vec{k}}/m_0 $, which is the inversion relation for $\alpha$-turbulence \citep{pierrehumbert_spectra_1994,smith_turbulent_2002,sukhatme_local_2009}. Figure \ref{F-mk_sigma} provides two examples, one with decreasing stratification (with $\alpha \approx 1.50$) and another with increasing stratification (with $\alpha \approx 0.49$). 
	
	To see how the parameter $\alpha$ modifies the resulting dynamics, consider a point vortex at the origin, given by $\xi =  \delta(\abs{\vec x})$, where $\abs{\vec x}$ is the horizontal distance from the vortex centre, and $\delta(\abs{\vec x})$ is the Dirac delta. If $\alpha=2$, then the streamfunction induced by the point vortex is logarithmic, $\psi(\abs{\vec x}) = \log(\abs{\vec x})/(2\pi)$. If $0 < \alpha < 2$, then $\psi(\abs{\vec x}) = -C_\alpha/\abs{\vec x}^{2-\alpha}$ where $C_\alpha>0$ is a constant \citep{iwayama_greens_2010}. Smaller $\alpha$ leads to  vortices with velocities decaying more quickly with the horizontal distance $\abs{\vec x}$, and hence a shorter interaction range. Thus, the vertical stratification modifies the relationship between a surface buoyancy anomaly and its induced velocity field: a surface buoyancy anomaly over decreasing stratification [$\sigma'(z)\leq 0$] generates a longer range velocity field than an identical buoyancy anomaly over increasing stratification [$\sigma'(z)\geq 0$].

	\subsection{Wave dispersion in variable stratification}
	
	The background gradient term, $\Lambda$, in the time-evolution equation \eqref{eq:time-evolution} allows for the propagation of surface-trapped Rossby waves. Substituting a wave solution of the form $\psi(x,z,t) =  \Psi_{k}(z) \, \exp \left[{\mathrm{i} \left(\vec k \cdot \vec r - \omega t \right)}\right]$, where the vertical structure $\Psi_k(z)$ satisfies the boundary value problem \eqref{eq:vertical_structure}\textendash \eqref{eq:lower}, into the time-evolution equation \eqref{eq:time-evolution} yields the angular frequency 
	\begin{equation}\label{eq:dispersion}
		\omega(\vec k) = - \frac{\Lambda\,k_x}{m(k)}.
	\end{equation}
	Given the relationship \eqref{eq:dBdy} between the  meridional surface buoyancy gradient $\mathrm{d}B/\mathrm{d}y|_{z=0}$ and the frequency $\Lambda$, a poleward decreasing buoyancy gradient ($f\mathrm{d}{B}/\mathrm{d}{y}<0$) implies westward propagating $(\omega<0)$ Rossby waves.
	
	The dispersion relation \eqref{eq:dispersion} shows that Rossby wave dispersion is coupled to the flow's interaction range and hence the stratification's vertical structure. If we approximate the inversion function as a power law \eqref{eq:mk_power} between $k_\mathrm{pyc} \lesssim k \lesssim k_s$, then the zonal phase speed, $c=\omega/k_x$, becomes $c\sim 1/k^\alpha$. Therefore, at these horizontal scales, Rossby waves are more dispersive over decreasing stratification (with $\alpha >1$) than over increasing stratification (with $\alpha <1$). In the limit that $\sigma_0 \gg \sigma_{\mathrm{pyc}}$ in which $\alpha \rightarrow 0$, then $c \approx $ constant, and so Rossby waves become non-dispersive.
	
	\section{From edge waves to surface-trapped jets}\label{S-staircase_theory}
	
	The emergence of jets in barotropic $\beta$-plane turbulence is due to two properties of the potential vorticity \citep{dritschel_multiple_2008,scott_zonal_2019}. The first is the resilience of strong latitudinal potential vorticity gradients to mixing \citep[i.e., "Rossby wave elasticity", ][]{dritschel_multiple_2008}. Regions with weak latitudinal potential vorticity gradients are preferentially mixed, weakening the gradient in these regions and enhancing the gradient in regions where the latitudinal potential vorticity gradient is already strong \citep{dritschel_jet_2011}. The ultimate limit of such latitudinally inhomogeneous mixing is a potential vorticity staircase \citep{danilov_barotropic_2004,dritschel_multiple_2008,scott_structure_2012}, which consists of uniform regions of potential vorticity punctuated by sharp potential vorticity gradients. The second property is that, through potential vorticity inversion, strong (positive) latitudinal gradients in potential vorticity correspond to eastward jets. Therefore, inverting a potential vorticity staircase produces a flow with eastward zonal jets centred at the sharp frontal zones, with weaker westward flows in between \citep{scott_zonal_2019}.

	%As discussed in \cite{dritschel_multiple_2008} and \cite{scott_zonal_2019}, two properties of potential vorticity lead to the emergence of jets in flows with a non-zero background meridional potential vorticity gradient, $\partial_y Q = \beta$.
	% The first property is that of Rossby wave elasticity, which describes the resilience of strong meridional potential vorticity gradients to latitudinal displacements; such displacements are dispersed as along-jet Rossby waves instead of turbulently mixing the background gradient. As a result, potential vorticity mixing predominantly takes place in regions where the meridional gradient is already weak. Under certain circumstances, this latitudinally inhomogeneous potential vorticity mixing transforms the background linear potential vorticity, $Q=\beta\,  y$, into a potential vorticity staircase.
	%  The second principle is that of potential vorticity inversion, which describes how the flow can be reconstructed from the potential vorticity field. Strong meridional gradients in the potential vorticity correspond to eastward jets. Thus, inverting a potential vorticity staircase produces a flow with eastward zonal jets centred at the sharp potential vorticity gradients, with westward return flows in between.
	
	However, the limit of a  potential vorticity staircase is only achieved for sufficiently large values of the non-dimensional number $\keps/\krh$ \citep{scott_structure_2012}, which is a ratio of the forcing intensity wavenumber, $\keps$, to the Rhines wavenumber, $\krh$. The forcing intensity wavenumber is given by \citep{maltrud_energy_1991}
	\begin{equation}
		\keps = (\beta^3/\varepsilon_{\mathcal{K}})^{1/5},
	\end{equation}
	 where $\varepsilon_{\mathcal{K}}$ is the kinetic energy injection rate in the barotropic model, and is obtained by setting the turbulent strain rate equal to the Rossby wave frequency \citep{vallis_generation_1993}. The Rhines wavenumber is given by \citep{rhines_waves_1975}
	 \begin{equation}\label{eq:rhines_2d}
		 \krh = \sqrt{\beta/U_\mathrm{rms}},
	 \end{equation}
	  where $U_\mathrm{rms}$ is the rms velocity. %, and  $1/\krh$ is proportional to the jet spacing \citep{dritschel_multiple_2008,dunkerton_barotropic_2008}. 
	 \cite{scott_structure_2012} found that the ratio $\keps/\krh$ controls the structure of zonal jets in barotropic $\beta$-plane turbulence; as $\keps/\krh$ is increased, the zonal jet strength increases and the potential vorticity gradient at the jet core becomes larger, with the staircase limit approached as $\keps/\krh \sim O(10)$.
		  
	 Jet formation in surface quasigeostrophic turbulence proceeds similarly, with the surface buoyancy (which is proportional to $\theta$) taking the role of the potential vorticity and the frequency, $\Lambda$, taking the role of the potential vorticity gradient, $\beta$. In this section, we first derive a non-dimensional number analogous to $\keps/\krh$ for surface quasigeostrophy. Then we consider how vertical stratification (and the non-locality parameter $\alpha$) modifies jet structure in the buoyancy staircase limit, as well as how it modifies the relationship between the Rhines wavenumber and the jet spacing.
	 
	 Before proceeding, we comment on two differences between two-dimensional barotropic turbulence and its surface quasigeostrophic counterpart. First, in the absence of forcing and dissipation, the kinetic energy,
	  \begin{equation}\label{eq:kinetic}
	  	\mathcal{K} = - \frac{1}{2}\, \overline{\psi \lap \psi } = \frac{1}{2} \, \overline{\abs{u}^2},
	  \end{equation}
	  is a conserved constant in two-dimensional barotropic turbulence (the overline denotes an area average). With a constant kinetic energy injection rate, $\varepsilon_\mathcal{K}$, and a linear damping rate, $r$, the equilibrium kinetic energy is $\mathcal{K}=\varepsilon_{\mathcal{K}}/2r$. By definition, the rms velocity is given by 
	    $U_\mathrm{rms} = \sqrt{2\mathcal{K}}$. Combining this expression with the definition of the kinetic energy \eqref{eq:kinetic} and substituting into the definition of the Rhines wavenumber \eqref{eq:rhines_2d} yields a Rhines wavenumber expressed in terms of external parameters alone,  
	    \begin{equation}
	    	\krh = \beta^{1/2} (r/\varepsilon_\mathcal{K})^{1/4}.
	    \end{equation}
	   In contrast, in surface quasigeostrophy, the total energy,
	   \begin{equation}
	  	\mathcal{E} = - \frac{1}{2} \overline{\psi|_{z=0}\, \theta },
	  \end{equation}
	  is a conserved constant in the absence of forcing and dissipation and there is no general relationship between the rms velocity, $U_\mathrm{rms}$, and the equilibrium total energy, $\mathcal{E}=\varepsilon/2r$, where $\varepsilon$ is the total energy injection rate in the surface quasigeostrophic model. Therefore, we are not generally able to express the Rhines wavenumber in terms of the external parameters $\varepsilon$, $\Lambda$, and $r$.
	  Second, because $\mathcal{E}$ and $\mathcal{K}$ have different dimensions, the kinetic energy injection in the barotropic model, $\varepsilon_{\mathcal{K}}$, has different dimensions than the total energy injection rate in the surface quasigeostrophic model, $\varepsilon$. In particular, $\varepsilon$ has dimensions of $L^2/T^3$.
	  
	  %length squared per time cubed.
	  
	% In contrast, in $\beta$-plane turbulence (with a vanishing deformation wavenumber), the kinetic energy, $\mathcal{K}$, is conserved  
	 % \begin{equation}
	 % 	\mathcal{K} =  \frac{1}{2} \overline{\abs{\grad \psi}^2_{z=0} }.
	 % \end{equation}
	%  Consequently, the energy injection rate, $\varepsilon$, in surface quasigeostrophy has different dimensions from its $\beta$-plane analogue, $\varepsilon_{\mathcal{K}}$. Second, in $\beta$-plane turbulence (with a vanishing deformation wavenumber), the equilibrium kinetic energy is given by $\mathcal{K} = \varepsilon_{\mathcal{K}}/2r$. Consequently, we can use $U = \sqrt{2\mathcal{K}}$ to express the Rhines wavenumber in terms external parameters alone, $\krh = \beta^{1/2} (r/\varepsilon)^{1/4}$. However, there is no analogous expression for the rms velocity, $U$, in surface quasigeostrophic turbulence.

	  \subsection{The forcing intensity wavenumber}

	  %To determine whether a certain latitudinal displacement  will disperse as Rossby waves or turbulently mix the background gradient, we can compare the Rossby wave frequency \eqref{eq:dispersion} to the turbulent strain rate, $\omega_s(k)$.
	  	  
	  To obtain the forcing intensity wavenumber, $\keps$, we compare the Rossby wave frequency \eqref{eq:dispersion} to the turbulent strain rate, $\omega_s(k)$.
	  If the inversion function is not approximately constant (i.e., $\alpha \neq 0$) then the strain rate is \citep{yassin_surface_2022}
	  \begin{equation}\label{eq:strain}
	  	\omega_s(k) \sim \varepsilon^{1/3}\, k^{4/3} \,\left[m(k)\right]^{-1/3}.
	  \end{equation}
	  In particular, if $m(k) = m_0 \,k^{\alpha}$, then $\omega_s(k) \sim m_0^{1/3}\varepsilon^{1/3}\, k^{\left(4-\alpha\right)/3}$. Setting the absolute value of the Rossby wave frequency for waves with $k = k_x$ equal to the turbulent strain rate \eqref{eq:strain} yields the condition
	  \begin{equation}\label{eq:keps_condition}
	  	\keps \left[m(\keps)\right]^{2} \sim \frac{\abs{\Lambda}^3}{\varepsilon}.
	  \end{equation}
	 A solution to this equation always exists because $\mathrm{d}m/\mathrm{d}k \geq 0$. If the inversion function takes the power law form \eqref{eq:mk_power}, then we obtain
	 \begin{equation}\label{eq:keps_alpha}
	 	\keps = \left(\frac{\abs{\Lambda}^3}{m_0^2\,\varepsilon}\right)^{1/\left(2\alpha+1\right)},
	 \end{equation}
	 which is equivalent to a wavenumber derived in \cite{smith_turbulent_2002}.

	\subsection{The damping rate wavenumber and the Rhines wavenumber}

	Suppose the inversion function takes an approximate power law form, $m(k) \approx m_0 \, k^\alpha$, near the energy containing wavenumbers.
    Then the generalization of the Rhines wavenumber at these wavenumbers is 
	\begin{equation}\label{eq:rhines}
		\krh = \left(\frac{\Lambda}{m_0 \, U_\mathrm{rms}}\right)^{1/\alpha}.
	\end{equation}
	 However, unlike in two-dimensional barotropic turbulence where $U_\mathrm{rms}= \sqrt{2\mathcal{K}} = \sqrt{\varepsilon_{\mathcal{K}}/r}$, we do not have a general relationship between $U_\mathrm{rms}$ and the external parameters $r$ and $\varepsilon$ in surface quasigeostrophic turbulence. To obtain a second wavenumber that depends on the damping rate, $r$, we follow \cite{smith_turbulent_2002}. From dimensional considerations, the energy spectrum at small wavenumbers is
	\begin{equation}\label{eq:Espectrum}
		E_\Lambda(k) \sim \Lambda^2 \, k^{-(\alpha+3)}/m_0.
	\end{equation}
	Then, defining $k_r$ as the wavenumber at which the inverse cascade halts, we obtain
	\begin{equation}
		\frac{\varepsilon}{2r} \approx  \int_{k_r}^\infty E(k) \, \mathrm{d}k \approx \left(\frac{\Lambda^2 /m_0}{\alpha+2}\right) k_r^{-(\alpha+2)},
	\end{equation}
	where the second equality follows because the integral is dominated by its peak at low wavenumbers. Solving for $k_r$ and neglecting any non-dimensional coefficients, we obtain
	\begin{equation}\label{eq:kr}
		k_r = \left(\frac{\Lambda^2 \, r}{m_0 \, \varepsilon}\right)^{1/(\alpha+2)}.
	\end{equation}
	%By definition, if the energy spectrum takes the form \eqref{eq:Espectrum}, then $k_r$ is the approximate wavenumber at which the inverse cascade is halted by the damping rate $r$. 
	Note that the damping rate wavenumber, $k_r$, has the same dependence on $\Lambda$, $\varepsilon$, and $r$ as the Rhines wavenumber, $\krh$, only if $\alpha=2$. 
	
	%agrees with the Rhines wavenumber only when $\alpha =2$.
			 
	 \subsection{Surface potential vorticity inversion} 
	
	A perfect surface potential vorticity staircase consists of mixed zones of halfwidth $b$, where $\mathrm{d}\theta/\mathrm{d}y  = -\Lambda$, separated by jump discontinuities at which $\mathrm{d}\theta/\mathrm{d}y = \infty$. We find it more conveniant to work with the relative surface potential vorticity, $\theta$, rather than the total surface potential vorticity, $\theta + \Lambda \, y$. In this case, if the total surface potential vorticity,  $\theta + \Lambda \, y$, is a perfect staircase with step width $2b$, then the relative surface potential vorticity, $\theta$, is a $2b$-periodic sawtooth wave. 
	
	Our first question is whether such a staircase is possible for general $m(k)$. To answer this question, we consider the velocity field induced by a jump discontinuity in $\theta$. For a jump discontinuity in an infinite domain,
	 \begin{equation}
		\theta =  
		\begin{cases}
			\Delta \theta  \quad  &\text{for} \quad 0< y< \infty\\
			0 \quad &\text{for} \quad  -\infty < y < 0 ,
		\end{cases}
	\end{equation}
	the zonal velocity is given by
	\begin{equation}
		u  = \frac{\Delta \theta}{2\pi} \int_{-\infty}^\infty \frac{\mathrm{e}^{\mathrm{i} \, k_y y}}{m\left(\abs{k_y}\right)} \mathrm{d}k_y.
	\end{equation}
	If $m(k)=m_0\,k^\alpha$, then this expression is proportional to $\abs{y}^{\alpha -1}$ if $\alpha \neq 1$ and logarithmic otherwise, and so the zonal velocity diverges at $y=0$ if $\alpha \leq 1$.  Consequently, we expect that a perfect staircase should not be possible over constant or increasing stratification due to the divergence of the zonal velocity at a jump discontinuity.

	%If $m(k) = m_0 \, k^\alpha$, this expression is proportional to $\abs{y}^{\alpha -1}$ if $\alpha \neq 1$. If $\alpha = 2$, $u$ is finite and continuous at $y=0$ but $\partial_y u$  has a jump discontinuity. Likewise, for $1 < \alpha < 2$, the zonal velocity is finite and continuous at $y=0$, but now the meridional gradient $\partial_y u$ suffers from an infinite discontinuity at $y=0$ because $\lim_{y\rightarrow 0^{\pm}} \partial_y u = \pm \infty$. Once $\alpha \leq 1$, the zonal velocity diverges at $y=0$, with a logarithmic divergence for $\alpha =1$ and a $\abs{y}^{\alpha -1}$ divergence for $\alpha <1$. Consequently, a perfect staircase should not be possible if $\alpha \leq 1$ (i.e., over constant or increasing stratification) due to the divergence of the zonal velocity.
	
	\begin{figure}
  \centerline{\includegraphics[width=0.9\columnwidth]{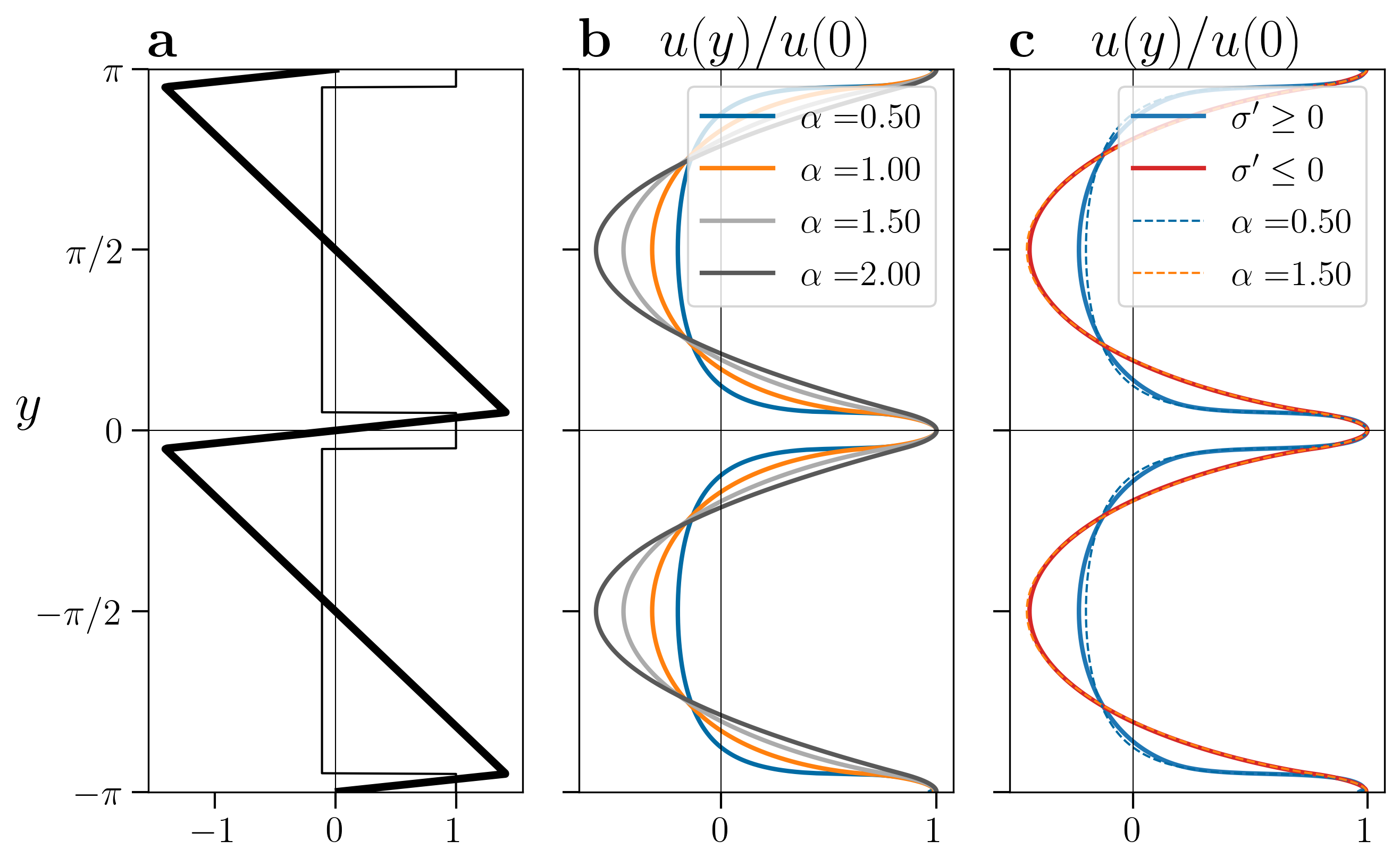}}
  \caption{Panel (a) shows a sloping sawtooth function (thick black line) along with its derivative (thin black line). Panel (b) shows the normalized zonal velocity induced by the sloping sawtooth function in panel (a) for various values of the parameter $\alpha$. Panel (c) shows the normalized zonal velocity induced by the sawtooth function in (a) in the increasing (blue line) and decreasing stratifications (red line) shown in figure \ref{F-mk_sigma}.}
  \label{F-sawtooth}
	\end{figure}
	
	We therefore consider the more general case of a sloping staircase, where there is a finite frontal zone of width $2a$ between the mixed zones. In this case, $\theta$ is a $2(a+b)$-periodic sloping sawtooth wave (see figure \ref{F-sawtooth}),  and is given by the periodic extension of
	\begin{equation}\label{eq:sloping_sawtooth}
		\theta = \Lambda
		\begin{cases}
			- \left[y-{(a+b)} \right] \quad &\text{for} \quad  {a}< y < {a+b} \\
			 \frac{\, b}{a} y \quad &\text{for} \quad \abs{y}\leq {a}\\
			- \left[y+ {(a+b)}\right] \quad &\text{for} \quad - {(a+b)}<y<- {a}.
		\end{cases}
	\end{equation}
	The meridional gradient $\mathrm{d}\theta/\mathrm{d}y$ is then a piecewise constant $2(a+b)$-periodic function
	\begin{equation}
		\d{\theta}{y} = \Lambda
		 \begin{cases}
			-1  \quad &\text{for} \quad  {a}< y < {a+b} \\
			 \frac{\, b}{a} \quad &\text{for} \quad \abs{y}\leq {a}\\
			-1  \quad &\text{for} \quad  -{(a+b)}<y<-{a}.
		\end{cases}
	\end{equation}
 	Therefore the gradient in the frontal zones exceeds the gradient in the mixed zones by a factor of $b/a$, which approaches infinity as $b/a \rightarrow \infty$ in the sawtooth wave limit.
 	
	The zonal velocity, $u=-\partial_y \psi$, is obtained by using the inversion relation \eqref{eq:inversion} to solve for the streamfunction. Alternatively, taking the meridional derivative of surface potential vorticity \eqref{eq:theta_inversion} gives
 	\begin{equation}
		\pd{\theta}{y} = \frac{1}{\sigma_0^2} \pd{u}{z}\Big|_{z=0}.
	\end{equation}
	Then in Fourier space [$\partial_y \rightarrow \mathrm{i} k_y$ and $\sigma_0^{-2}\partial_z|_{z=0} \rightarrow m(k)$] we obtain
	\begin{equation}
		\hat u_{\vec k} = \frac{1}{m(k)} \left( \mathrm{i}\,  k_y \, \hat \theta_{\vec k} \right),
	\end{equation}
	which shows that the induced zonal velocity is obtained by smoothing $\mathrm{d}\theta/\mathrm{d}y$ by the function $m(k)$. An immediate consequence is that the east-west asymmetry in the zonal velocity is fundamentally due to the east-west asymmetry in the gradient $\mathrm{d}\theta/\mathrm{d}y$.

\begin{figure}
  \centerline{\includegraphics[width=0.75\columnwidth]{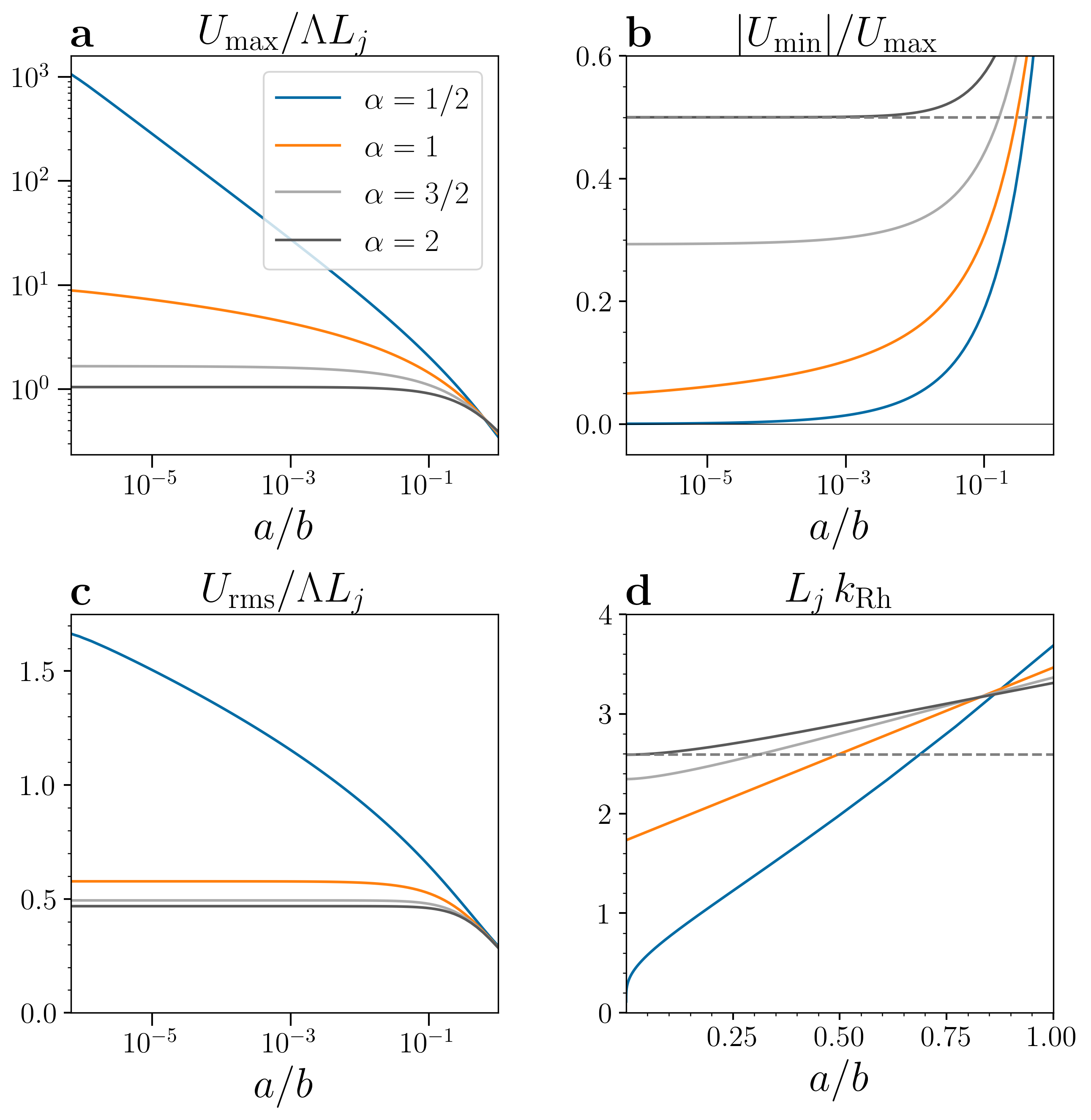}}
  \caption{Properties of zonal velocity profiles induced by sloping sawtooth profiles \eqref{eq:sloping_sawtooth} of $\theta$ as a function of the non-dimensional frontal zone width $a/b$ separating the mixed zones for four values of $\alpha$. Panel (a) shows the maximum zonal velocity, panel (b) shows the ratio of westward speed to eastward speed, panel (c) shows the rms zonal velocity, and panel (d) shows the product $L_j \krh$ where $L_j=a+b$ is the halfwidth separation (the distance between $U_\mathrm{mix}$ and $U_\mathrm{min}$) and $\krh$ is the Rhines wavenumber \eqref{eq:rhines}. }
  \label{F-velocity_sawtooth}
\end{figure}

	Figure \ref{F-sawtooth} shows an example of sloping sawtooth $\theta$ profile along with the induced zonal velocities. 
	For a power law inversion function, $m(k) = m_0 k^{\alpha}$, the parameter $\alpha$ modifies the zonal velocity in two ways. 
	First, in more local flows (with smaller $\alpha$), the zonal velocity decays more rapidly away from the jet centre, as expected. 
	Second, the degree of smoothing increases with $\alpha$, and so more local regimes (with smaller $\alpha$) are more east-west asymmetric, with the ratio $\abs{u_\mathrm{min}}/u_\mathrm{max}$ taking smaller values for smaller $\alpha$.
	Figure \ref{F-velocity_sawtooth}(b) shows $\abs{u_\mathrm{min}}/u_\mathrm{max}$ as a function of $a/b$ for $\alpha \in \{1/2,\,1,\,3/2,\,2\}$. For $\alpha = 2$, we obtain $\abs{u_\mathrm{min}}/u_\mathrm{max} \rightarrow 1/2$ in the limit $a/b \rightarrow 0$ so that eastward jets are only twice as strong as westward flows in the perfect staircase limit \citep{danilov_scaling_2004,dritschel_multiple_2008}. At $\alpha = 3/2$, we find $\abs{u_\mathrm{min}}/u_\mathrm{max} \approx 0.29$ in the $a/b \rightarrow 0$ limit so that eastward jets are now more than three time as strong as westward flows. Once $\alpha \leq 1$, then the maximum jet velocity diverges as $\alpha \rightarrow 0$ [figure \ref{F-velocity_sawtooth}(a)] and so $\abs{u_\mathrm{min}}/u_\mathrm{max} \rightarrow 0$ as $a/b \rightarrow 0$.
	
	If $m(k)$ is not a power law, then the results are similar so long as $m(k)$ can be approximated by a power law at small wavenumbers. 
	Figure \ref{F-sawtooth} shows the induced velocity for the inversion functions computed from idealized stratifications profiles (shown in figure \ref{F-mk_sigma}). Because these inversion functions can be approximated by power laws $m(k) \approx k^{0.49}$ and $m(k) \approx k^{1.50}$ at small wavenumbers, the induced velocity fields nearly coincide with the velocity fields computed from power law inversion functions with $\alpha =0.5$ and $\alpha = 1.5$.
	
 	Finally, we examine how the Rhines wavenumber, $\krh$, relates to jet spacing. Let
 	\begin{equation}
 			 L_j=a+b 
 	\end{equation}
	 be the half-separation between the jets, i.e., the half distance between consecutive zonal velocity maxima.
	For two-dimensional barotropic turbulence (i.e., the $\alpha = 2$ case), we have $L_j = 45^{1/4}/\krh \approx 2.59/\krh$
 	in the staircase limit \citep[i.e, for $a/b \rightarrow 0$,][]{dritschel_multiple_2008,scott_structure_2012}. This result is found by solving for the zonal velocity induced by a staircase with halfwidth $L_j=b$, taking the rms of the zonal velocity, and then substituting into the definition of the generalized Rhines wavenumber \eqref{eq:rhines}. As figure \ref{F-velocity_sawtooth}(d) shows, because the velocity field induced by a perfect staircase depends on the inversion function, $m(k)$, the relationship between $L_j$ and $\krh$ also depends on the inversion function. For $m(k) = k^{3/2}$, an analogous calculation gives $L_j  \approx 2.35/\krh$ in the staircase limit. For $\alpha = 1$, even though the maximum velocity diverges at $a/b \rightarrow 0$, the rms velocity asymptotes to a constant value, and so we obtain a half jet-separation of $L_j  \approx 1.73/\krh$ (figure \ref{F-velocity_sawtooth}). Finally in the $\alpha = 1/2$ case, although the rms speed has not converged by $a/b = 10^{-6}$, the product $L_j \, \krh$ is approaching values close to zero.
 		 
\section{Numerical Simulations}\label{S-numerical}
	
	\subsection{The numerical model}
	
	We use the \texttt{pyqg} pseudo-spectral model \citep{abernathey_pyqgpyqg_2019} which solves the time-evolution equation \eqref{eq:time-evolution} in a square domain with side length $L=2\pi$.  Time-stepping is through a third-order Adam-Bashforth scheme with small-scale dissipation achieved through a scale-selective exponential filter \citep{smith_turbulent_2002,arbic_coherent_2003},
	\begin{equation}
		\mathrm{ssd} =
		 \begin{cases}
			1 \quad \text{for} \quad  k \leq k_0 \\
			e^{-a(k-k_0)^4} \quad  \text{for} \quad k > k_0,
		\end{cases}
	\end{equation}
	with $a = 23.6$ and $k_0 = 0.65 k_\mathrm{Nyq}$ where $k_\mathrm{Nyq}=\pi$ is the Nyquist wavenumber. The forcing is isotropic, centred at wavenumber $k_f = 80$, and normalized so that the energy injection rate is $\varepsilon = 1$ \citep[see appendix B in][]{smith_turbulent_2002}. However, the effective energy injection rate, $\varepsilon_\mathrm{eff}$, is smaller than $\varepsilon$ due to dissipation.  To determine $\varepsilon_\mathrm{eff}$ from numerical simulations, we use $ \varepsilon_\mathrm{eff}= 2\, r \, \mathcal{E}$ where $\mathcal{E}$ is the equilibrated total energy diagnosed from the model. In what follows, we report values of $\keps/k_r$ using $\varepsilon_\mathrm{eff}$ instead of $\varepsilon$. The model is integrated forward in time until at least  $t=5/r$ to allow the fluid to reach equilibrium. All model runs use $1024^2$ horizontal grid points. 
	
	\subsection{For what values of $\keps/\kr$ do jets form?}\label{SS-sims}
	
	  \begin{figure}
  \centerline{\includegraphics[width=1\columnwidth]{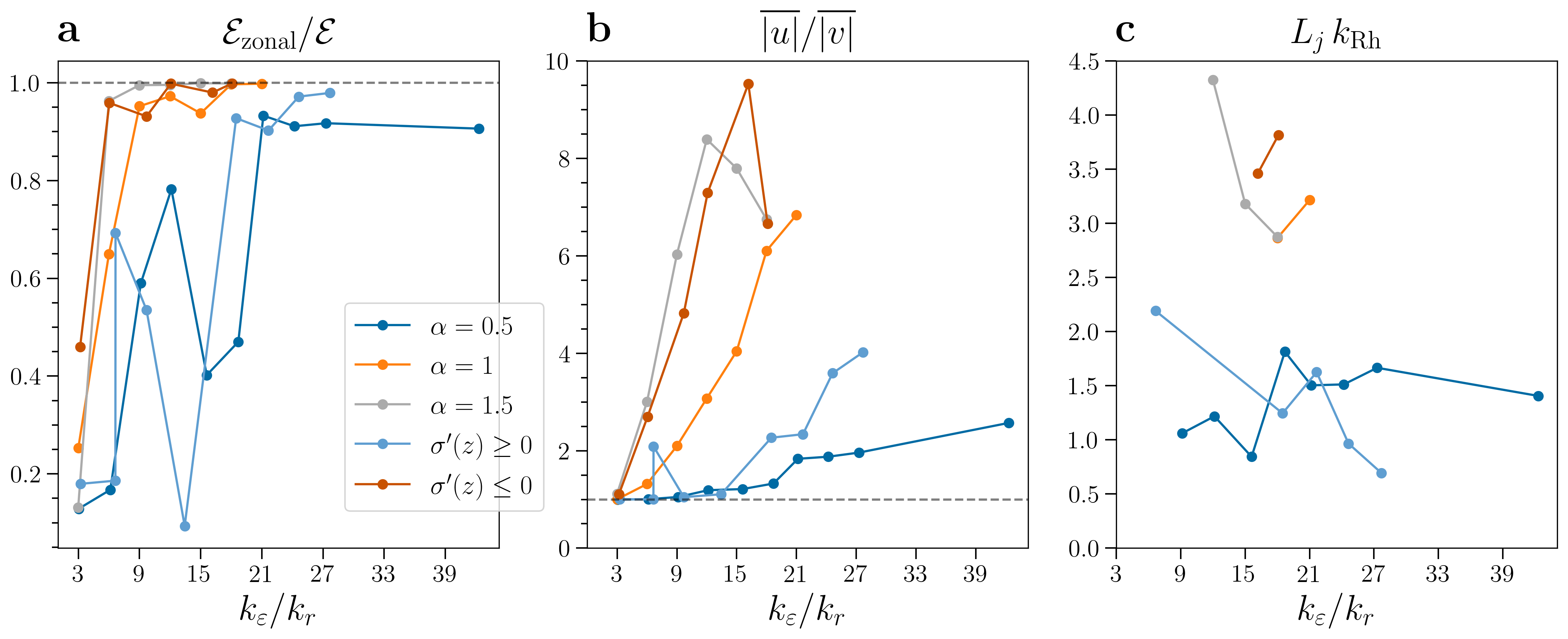}}
  \caption{Diagnostics from five series of simulations as a function of the non-dimensional number $\keps/k_r$. The first three series of simulations have inversion function $m(k)=k^\alpha$ with $\alpha \in \{\nicefrac{1}{2},\, 1,\, \nicefrac{3}{2}\}$. For the other two series, the inversion functions are shown in figure \ref{F-mk_sigma_alt}. Panel (a) shows the ratio of energy in the zonal mode to total energy. Panel (b) shows the ratio of domain averaged zonal speed to domain averaged meridional speed. Panel (c) shows the ratio of westward zonal speed to eastward zonal speed. Panel (d) shows the relationship between the halfwidth jet spacing, $L_j$, and the Rhines wavenumber, $\krh$.}
  \label{F-krh_velocity}
   \end{figure}
	
	For our first set of simulations, we vary $\keps/k_r$ over the values shown in figure \ref{F-krh_velocity}. We do so by fixing $k_r=8$ and varying $\keps$. For a given value of $\keps$, we choose $\Lambda$ and $r$ so as to maintain $\kr = 8$ (the energy injection rate, $\varepsilon$, is fixed at unity for all model runs). Given $\kr$ and $\keps$, we rearrange the definition of $k_r$ \eqref{eq:kr} to solve for $\gamma= r \, \Lambda^2$,
	\begin{equation}
		\gamma = m_0 \, \varepsilon \, k_r^{\alpha+2},
	\end{equation}
	then solve for $r$ in the implicit equation \eqref{eq:keps_condition} for $\keps$ ,
	\begin{equation}
		 r = \frac{\gamma}{\left( \varepsilon \, \keps \, [m(\keps)]^2 \right)^{2/3}},
	\end{equation}
	and finally use the definition $\gamma = r\, \Lambda^2$ to solve for $\Lambda$.

	%We run five series of simulations, each with a different prescribed $m(k)$. In three of these series, we specify $m(k) = k^\alpha$ for $\alpha \in \{1/2, 1, 3/2 \}$.
	 %For the other two series, we specify a stratification $\sigma(z)$ of the form given by equation \eqref{eq:sigma_piece} and then find $m(k)$ by solving the boundary value problem \eqref{eq:vertical_structure}\textendash \eqref{eq:lower} at each model wavenumber. The stratification profiles and the resulting inversion functions are shown in figure \ref{F-mk_sigma_alt}.
	 %One case consists of a non-decreasing stratification  profile [$\sigma^\prime(z) \geq 0$] with $\sigma_0 = 1$, $\sigma_\mathrm{pyc} = 0.1$, $h_\mathrm{mix} =0.01$ and $h_\mathrm{lin}=0.05$. The resulting $m(k)$ is approximately linear for $k \gtrsim 70$ and transitions to an approximate sub-linear wavenumber dependence $m(k) \sim k^{0.40}$ for wavenumbers $5 \lesssim k \lesssim 50$. The second case consists of a non-increasing stratification profile [$\sigma^\prime(z) \leq 0$] with $\sigma_0 = 0.13$, $\sigma_\mathrm{pyc} = 1$, $h_\mathrm{mix} =0.125$ and $h_\mathrm{lin}=0.2$. The resulting $m(k)$ is approximately linear at wavenumbers $k \gtrsim 60$ and transitions to an approximate super linear wavenumber dependence $m(k) \sim k^{1.50}$ between $3 \lesssim k \lesssim 60$.
	 
	\subsubsection{Power law inversion functions}
	We first describe the results from three series of simulations with power law inversion functions, $m(k)=k^\alpha$, with $\alpha \in \{\nicefrac{1}{2},\, 1, \, \nicefrac{3}{2} \}$.
	Summary diagnostics from these simulations are shown in figure \ref{F-krh_velocity}.	
	In panel (a), we observe that the ratio of energy in the zonal mode to total energy, $\mathcal{E}_\mathrm{zonal}/\mathcal{E}$, increases with $\keps/k_r$, and that the majority of the total energy is in the zonal mode for sufficiently large $\keps/k_r$. For a fixed $\keps/k_r$, more of the total energy is zonal in more non-local flows (with larger $\alpha$) than in more local flows (with smaller $\alpha$); for $\alpha = \nicefrac{3}{2}$, we have $\mathcal{E}_\mathrm{zonal}/\mathcal{E} \approx 1$ by $\keps/k_r \approx 6$ as compared to $\keps/k_r \approx 12$ for $\alpha = 1$. Moreover, for $\alpha = \nicefrac{1}{2}$, we find that $\mathcal{E}_\mathrm{zonal}/\mathcal{E}$ asymptotes to approximately 0.9 once $\keps/k_r \approx 18$ with little subsequent change for larger values of $\keps/k_r$. In panel (b), we observe a striking contrast in the ratio $\overline{\abs{u}}/\overline{\abs{v}}$ between different values of $\alpha$ (the overline denotes a domain average). For $\alpha = \nicefrac{3}{2}$, the domain averaged zonal speed, $\overline{\abs{u}}$, is approximately eight times larger than the domain averaged  meridional speed, $\overline{\abs{v}}$, for large $\keps/k_r$.  In contrast, for $\alpha = \nicefrac{1}{2}$, $\overline{\abs{u}}$ only exceeds  $\overline{\abs{v}}$ by a multiple of two for large $\keps/k_r$.
		
	\begin{figure}
  \centerline{\includegraphics[width=.8\columnwidth]{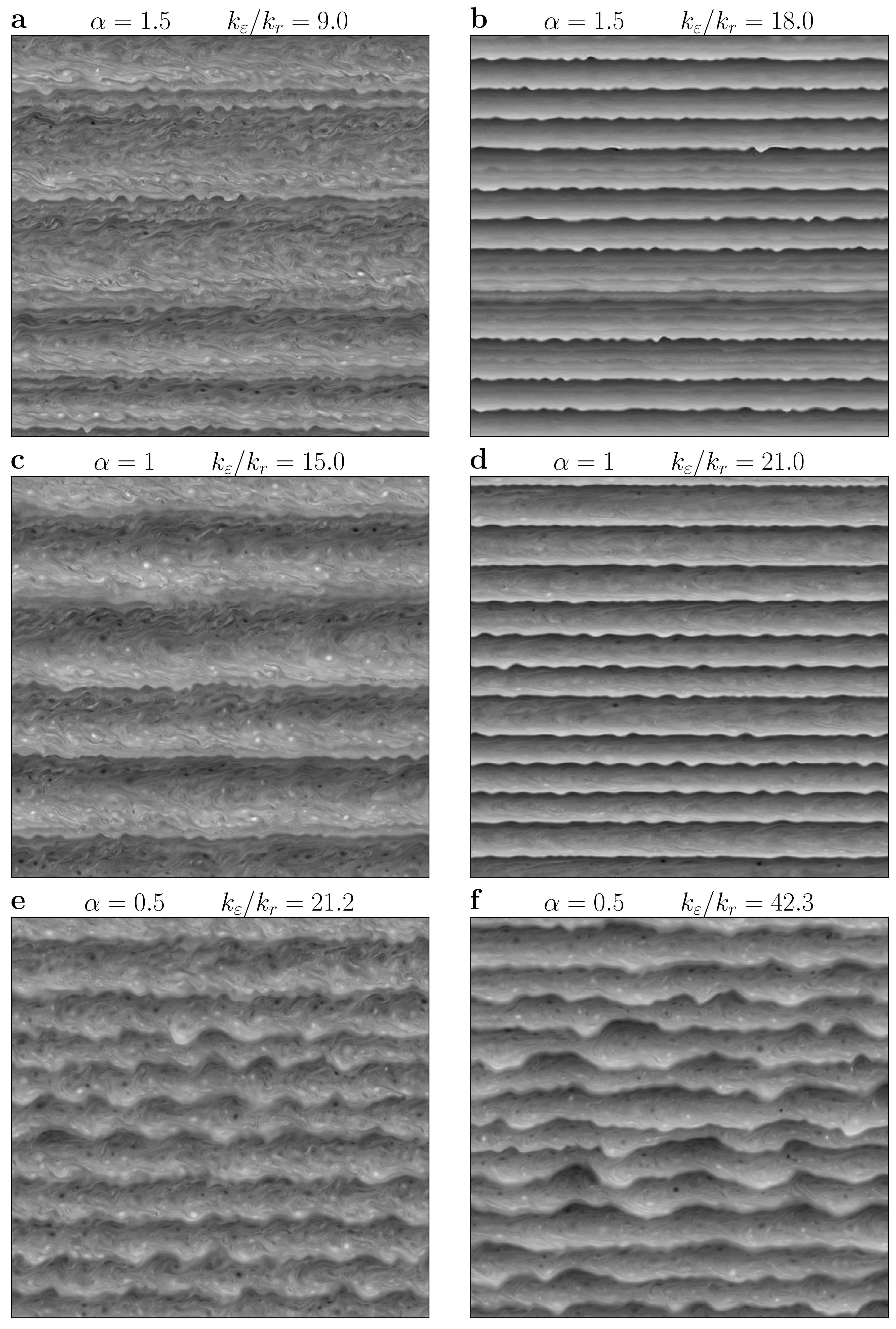}}
  \caption{Snapshots of the relative surface potential vorticity, $\theta$, for simulations with power law inversion functions, $m(k)=k^\alpha$. In each snapshot, the $\theta$ field is normalized by its maximum value in the snapshot. Only one quarter of the domain is shown (i.e., $512^2$ grid points). }
  \label{F-alpha_jets}
   \end{figure}
   
   	\begin{figure}
  \centerline{\includegraphics[width=1\columnwidth]{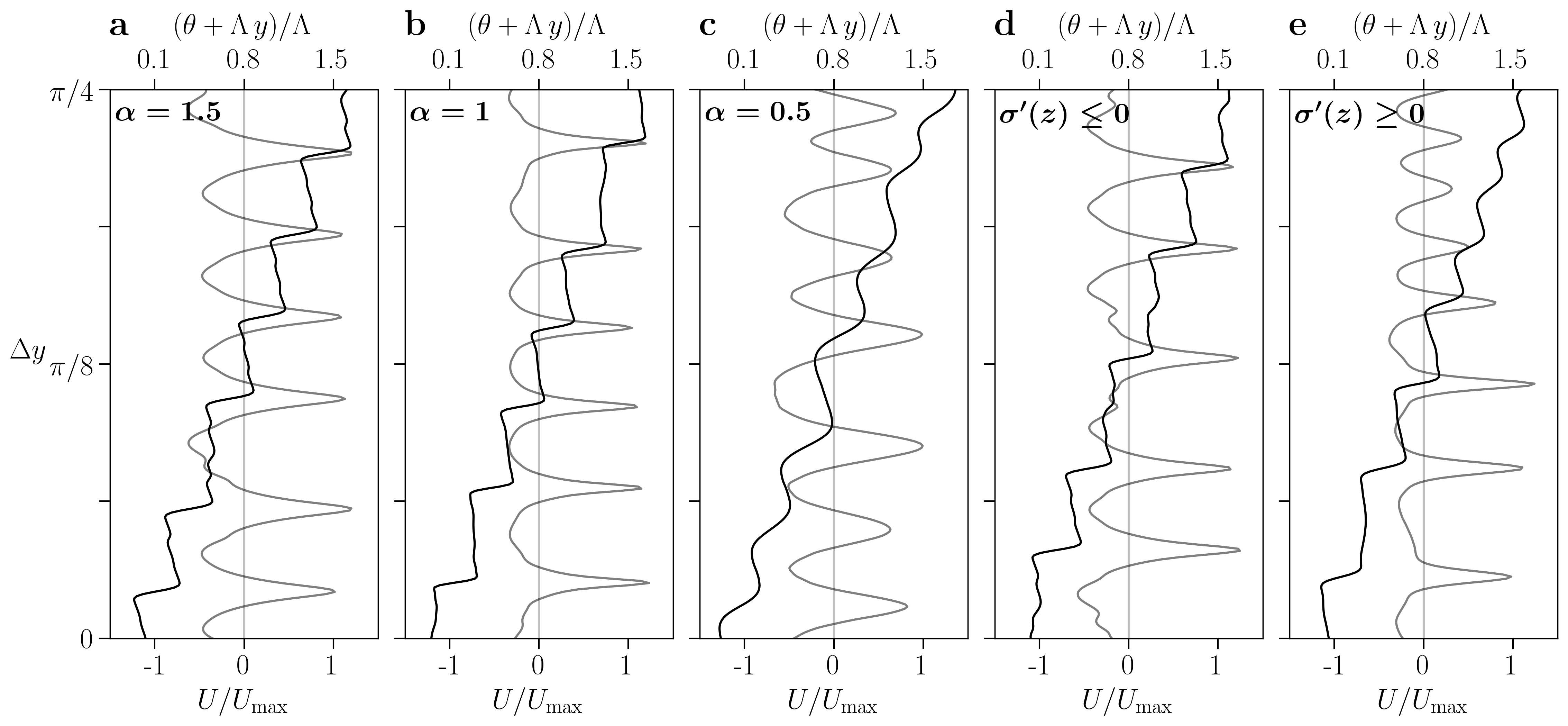}}
  \caption{The zonal mean total surface potential vorticity, $\theta + \Lambda \, y$, in black and the zonal mean zonal velocity, $U$, in grey. }
  \label{F-model_staircase}
	\end{figure}
 
	Next, we examine the jet structure for different $\alpha$ as a function of $\keps/k_r$. Figure \ref{F-alpha_jets} shows $\theta$-snapshots from model runs with $m(k)=k^\alpha$. For each value of $\alpha$, two model runs are shown: one where jets have just become visible in the $\theta$-snapshot and another with the largest value of $\keps/k_r$, which we expect to be  closest to the staircase limit. The jets are visible in these snapshots as the regions with strong gradients. Because these are $\theta$-snapshots rather than $(\theta+\Lambda\,y)$-snapshots, the $(\theta + \Lambda \, y)$-staircase is instead a $\theta$-sawtooth, and the mixed zones between the jets are approximately linear in $\theta$. 
	%Indeed, in all cases, the mixed zones separating the sharp jumps appear approximately linear in the model runs with large $\keps/k_r$.
	 We confirm this to be the case in figure \ref{F-model_staircase}, where the zonal averages of the total surface potential vorticity, $\theta+\Lambda \, y$, and the zonal velocity are shown. For the  $\alpha = \nicefrac{3}{2}$ and $\alpha =1$ cases, we observe an approximate staircase structure with nearly uniform mixed zones separated by frontal zones, and with jets centred at sharp $\theta$ gradients. As expected from the idealized staircases of section \ref{S-staircase_theory}, close to the staircase limit, the $\alpha=1$ jets are narrower than the $\alpha=\nicefrac{3}{2}$ jets, and the ratio of maximum westward speed to maximum eastward speed, $|U_\mathrm{min}|/|U_\mathrm{max}|$, is smaller at $\alpha=1$ than at $\alpha=\nicefrac{3}{2}$.
	 
	 %As expected, the $\alpha =1$ jets are narrower than the $\alpha = 3/2$ jets.
	 %In addition the ratio $U_\mathrm{min}/U_\mathrm{max}$ is smaller for $\alpha=1$ jets than for $\alpha = 3/2$ jets, as expected from the idealized staircases of section \ref{S-staircase_theory}.	 

	 In contrast to the $\alpha = \nicefrac{3}{2}$ and the $\alpha = 1$ series, the $\alpha=\nicefrac{1}{2}$ series approaches the staircase limit slowly with $\keps/k_r$. 
	The $\alpha=\nicefrac{1}{2}$ staircase remains smooth even at $\keps/k_r = 42$ [figure \ref{F-model_staircase}(c)]. The ratio of frontal zone width to mixed zone width, $a/b$, is between $0.5$ and  $0.65$ for $\alpha=\nicefrac{1}{2}$ jets. In contrast, this ratio is between $0.15$ and $0.2$ for the $\alpha =\nicefrac{3}{2}$ and $\alpha = 1$ jets.
	 In part, the broadness of the $\alpha=\nicefrac{1}{2}$ frontal zones is a consequence of zonal averaging in the presence of large amplitude undulations. However, it is evident from the $\theta$-snapshots of figure \ref{F-alpha_jets} that the $\alpha=\nicefrac{1}{2}$ frontal zones are indeed broader than the $\alpha=\nicefrac{3}{2}$ and $\alpha=1$ frontal zones [e.g., compare panels (a) and (d) with (f) in figure \ref{F-alpha_jets}], even without zonal averaging. 
	 %Moreover, from figure \ref{F-krh_velocity}(a), the ratio of zonal energy to total energy, $\mathcal{E}_\mathrm{zonal}/\mathcal{E}$, also asymptotes to a value of $0.9$, with little change in the ratio once $\keps/k_r$ is exceeds 18.
	 
	%  do not reach the staircase limit, even for the largest values of $\keps/k_r$. Instead, we obtain a smoothed staircase, and, instead of sharp discontinuities separating the mixed zones, there is a relatively wide ($\approx 0.6$ of the mixed zone width) region with a positive gradient, leading to broad jets [figure \ref{F-model_staircase}(c)].
	% The broadness of the $\alpha=1/2$ jets can also be verified by comparing the $\alpha=1/2$ jets with the  $\alpha=3/2$ or $\alpha=1$ jets in the $\theta$ snapshots of figure \ref{F-alpha_jets} for the larger values of $\keps/k_r$.  
	
	 We now examine how the generalized Rhines wavenumber, $\krh$, relates to the jet spacing.
	 %The inverse Rhines wavenumber is proportional the jet spacing, with the constant of proportionality depending on $\alpha$.
	 From figure \ref{F-velocity_sawtooth}(d), a ratio of $a/b \approx 0.2$ leads to a $L_j \, \krh \approx 2.2$ for $\alpha=\nicefrac{3}{2}$ and $L_j \, \krh \approx 2.0$ for $\alpha=1$. But as figure \ref{F-krh_velocity}(d) shows, we find values closer to $L_j \, \krh \approx 3$ for both of these cases. In contrast, for the $\alpha =\nicefrac{1}{2}$ jets, figure \ref{F-velocity_sawtooth}(d) predicts $1.98 \lesssim L_j \, \krh \lesssim 2.5$ for the observed range of $0.5 \lesssim a/b \lesssim 0.65$, but we find $L_j \, \krh \approx 1.5$ for $\keps/ k_r \geq 18$, which is smaller than predicted.  
	 
	 %The idealized staircases of section \ref{S-staircase_theory} indicate that $b\, \krh$ is smaller for $\alpha=1$ than for $\alpha = 3/2$. However, the values diagnosed from the model are fairly similar, with $b \, \krh$ between 2.5 and 3.5 [figure \ref{F-krh_velocity}(d)]. A perfect staircase, in contrast, has $b\,\krh \approx 2.35$ for $\alpha=3/2$ and  $b\,\krh \approx 1.73$ for $\alpha=1$. In the $\alpha=0.5$ case, we obtain $b\, \krh \approx 1$ for larger values of $\keps/k_r$.
	 % In both cases, this is a consequnce of the finite $a/b$.........
	 	 
	Returning to figure \ref{F-alpha_jets}, we observe that there are undulations along the jets, with smaller values of $\alpha$ corresponding to larger amplitude undulations. 
	 These undulations propagate as waves and are less dispersive for smaller $\alpha$, propagating eastward for $\alpha=\frac{3}{2}$, westward for $\alpha=\nicefrac{1}{2}$,  and are nearly stationary for $\alpha = 1$. Moreover, the waves in the $\alpha=\nicefrac{1}{2}$ case maintain their shape as they propagate for a significant fraction of the domain, although they eventually disperse or merge with other along jet waves. 
	 That we obtain larger amplitude along jet undulations for smaller $\alpha$ is a consequence of the more local inversion operator \eqref{eq:inversion} at smaller $\alpha$. A jet in a highly local flow (with small $\alpha$) is ``a coherent structure that hangs together strongly while being easy to push sideways'' \citep[][in the context of equivalent barotropic jets]{mcintyre_potential-vorticity_2008}. However, although both an equivalent barotropic jet and an $\alpha=\nicefrac{1}{2}$ jet exhibit large meridional undulations, the undulations in the equivalent barotropic case are frozen in place \citep[because of a vanishing group velocity at large scales, ][]{mcintyre_potential-vorticity_2008} and so the equivalent barotropic jet behaves like a meandering river with a fixed shape. In contrast, the $\alpha=\nicefrac{1}{2}$ jet behaves like a flexible string whose shape evolves in time with the propagation of weakly dispersive waves. Another difference between the two cases is that an equivalent barotropic jet has a width given by the deformation radius. In contrast, there is no analogous characteristic scale for $\alpha=\nicefrac{1}{2}$ jets and, in principle, the jets should become infinitely thin as $\keps/k_r \rightarrow \infty$.
	 	 	
	\begin{figure}
  \centerline{\includegraphics[width=.9\columnwidth]{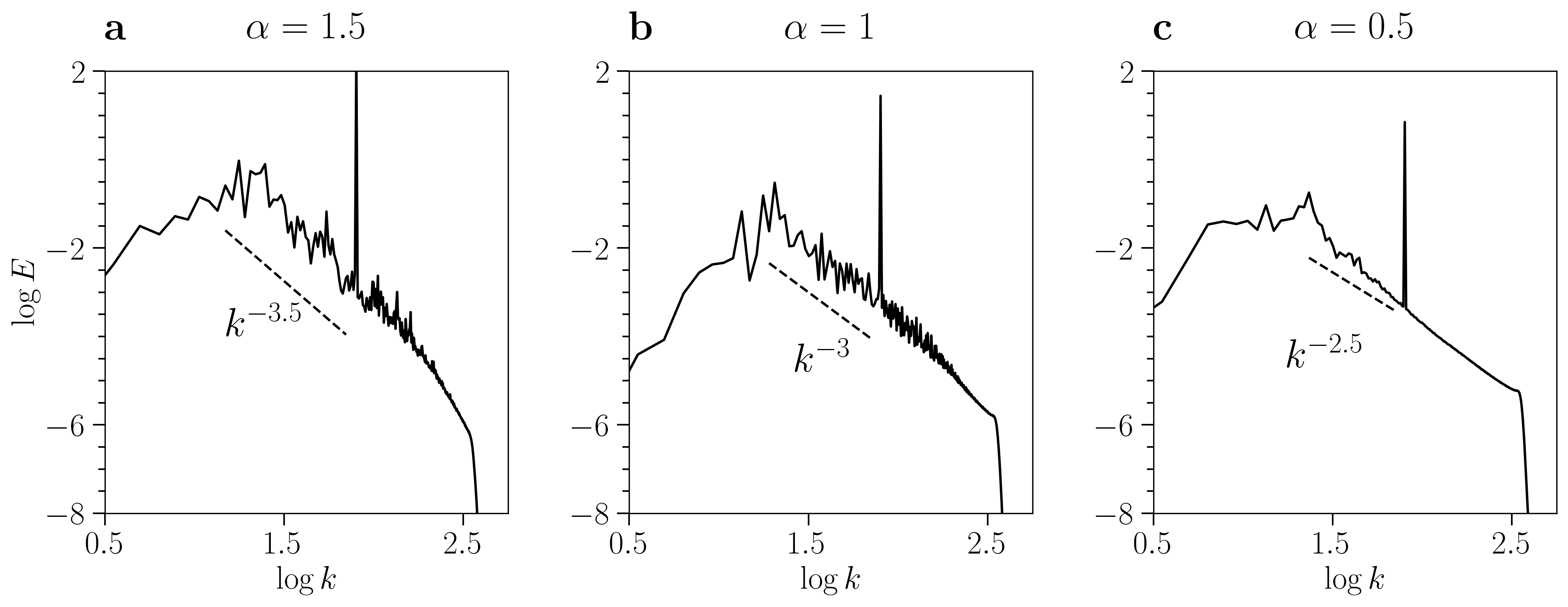}}
  \caption{The total energy spectrum, $E(k)$, as a function of the wavenumber, $k=k_x^2+k_x^2$, for three simulations with power law inversion functions, $m(k)=k^\alpha$. The values of $\keps/k_r$ are 18.0 for panel (a), 21.0 for panel (b), and 42.3 for panel (c).}
  \label{F-energy_spectrum}
	\end{figure}

	Energy spectra for the three power law simulations are shown in figure \ref{F-energy_spectrum}. The energy spectrum obtained from dimensional analysis  \eqref{eq:Espectrum} gives a $k^{-\alpha - 3}$ wavenumber dependence, which leads to the familiar $k^{-5}$ spectrum for beta-plane barotropic turbulence ($\alpha=2$). Although early investigations \citep{chekhlov_effect_1996,huang_anisotropic_2000,danilov_barotropic_2004} found a $k^{-5}$ spectrum in barotropic $\beta$-plane turbulence, \cite{scott_structure_2012} instead found a shallower $k^{-4}$ spectrum in the staircase limit \citep[suggested earlier by][]{danilov_barotropic_2004,danilov_scaling_2004}, which they explained as a consequence of the sharp discontinuities of the staircase. Generalizing their argument to the present case, a one dimensional $\theta(y)$ series with discontinuities implies a Fourier series with coefficients decaying as $k^{-1}$, leading to a $\theta^2$ spectrum of $k^{-2}$, and hence an energy spectrum 
	\begin{equation}
		E(k) \sim k^{-2}\left[m(k)\right]^{-1}. %\frac{1}{k^2\,m(k)}. 
	\end{equation}
	If $m(k) \sim k^{\alpha}$, then we obtain a spectrum $E(k) \sim k^{-\alpha - 2}$, which yields the $k^{-4}$ spectrum observed in \cite{scott_structure_2012}, where $\alpha=2$. For $\alpha=\nicefrac{3}{2}$, $\alpha=1$, and $\alpha=\nicefrac{1}{2}$, the predicted spectrum is proportional to $k^{-3.5}$, $k^{-3}$, and $k^{-2.5}$, respectively.
	The diagnosed spectra shown in figure \ref{F-energy_spectrum} are consistent with these shallow spectra, instead of energy spectrum  \eqref{eq:Espectrum} obtained from dimensional considerations.

	\subsubsection{Inversion functions from $\sigma(z)$}
	
	\begin{figure}
  \centerline{\includegraphics[width=.7\columnwidth]{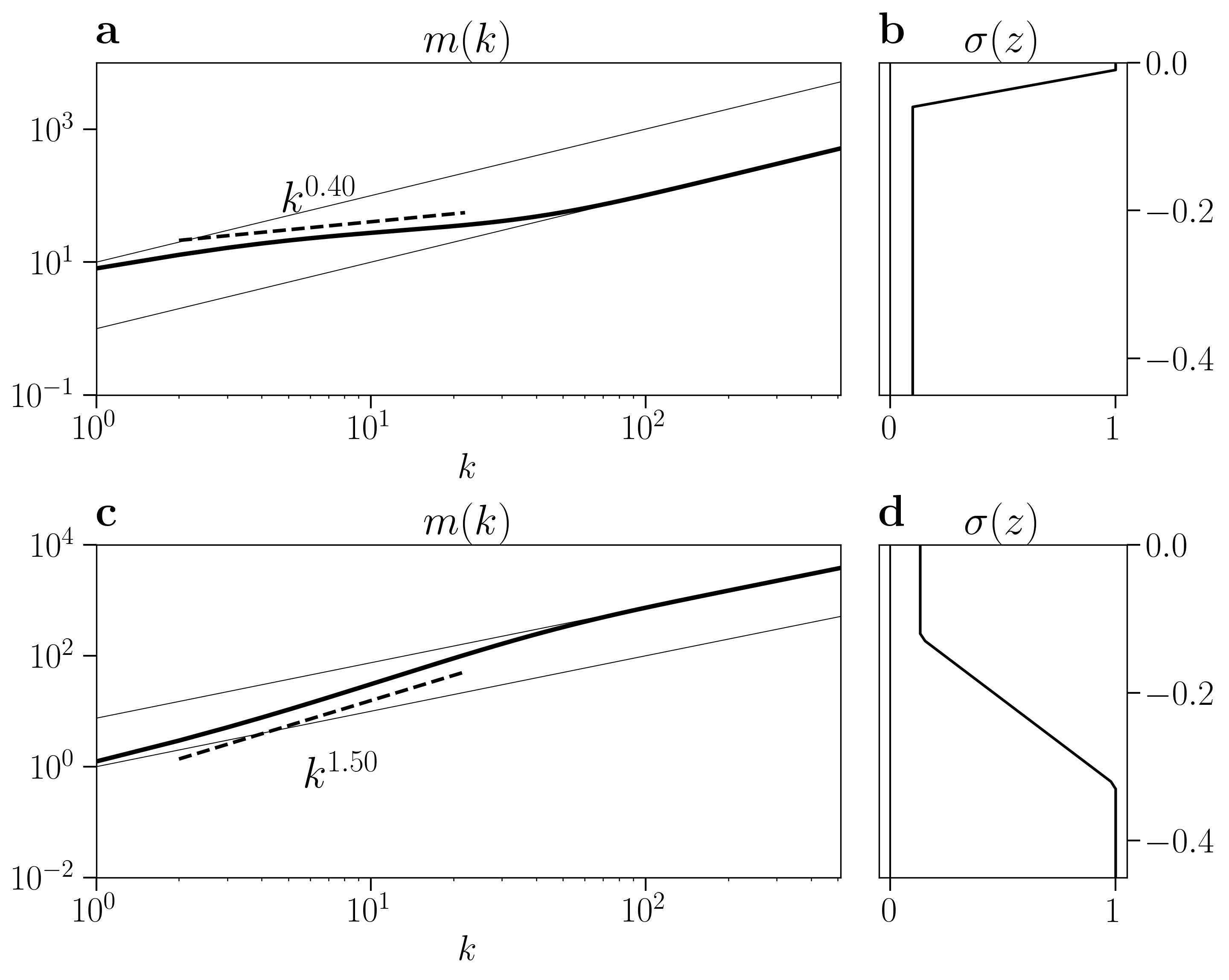}}
  \caption{Inversion functions  [panel (a) and (c)] along with their corresponding stratification profiles [panels (b) and (d), respectively]. The stratification profiles are given by the piecewise function \eqref{eq:sigma_piece}. For panel (a), we have $\sigma_0 = 1$, $\sigma_\mathrm{pyc}=0.1$, $h_\mathrm{mix}=0.01$, and $h_\mathrm{lin}=0.05$. For panel (c), we have $\sigma_0 = 0.133$, $\sigma_\mathrm{pyc}=1$, $h_\mathrm{mix}= 0.125$, and $h_\mathrm{lin}=0.2$. The thin grey lines in panels (a) and (c) are given by $k/\sigma_0$ and $k/\sigma_\mathrm{pyc}$.   }
  \label{F-mk_sigma_alt}
   \end{figure}
   
   We also ran two series of simulations where we specified a piecewise stratification profile \eqref{eq:sigma_piece}, and then obtained $m(k)$ by solving the boundary value problem \eqref{eq:vertical_structure}\textendash \eqref{eq:lower} at each wavenumber. The stratification profiles and the resulting inversion functions are shown in figure \ref{F-mk_sigma_alt}.
	 One case consists of an increasing stratification  profile [$\sigma^\prime(z) \geq 0$] with $\sigma_0 = 1$, $\sigma_\mathrm{pyc} = 0.1$, $h_\mathrm{mix} =0.01$ and $h_\mathrm{lin}=0.05$. The resulting $m(k)$ is approximately linear for $k \gtrsim 70$ and transitions to an approximate sub-linear wavenumber dependence $m(k) \sim k^{0.40}$ for wavenumbers $5 \lesssim k \lesssim 50$. The second case consists of a decreasing stratification profile [$\sigma^\prime(z) \leq 0$] with $\sigma_0 = 0.13$, $\sigma_\mathrm{pyc} = 1$, $h_\mathrm{mix} =0.125$ and $h_\mathrm{lin}=0.2$. The resulting $m(k)$ is approximately linear at wavenumbers $k \gtrsim 60$ and transitions to an approximate super linear wavenumber dependence $m(k) \sim k^{1.50}$ between $3 \lesssim k \lesssim 60$.
	 
	As seen in figure \ref{F-krh_velocity}, the $\sigma'(z) \leq 0$ case is similar to the $\alpha=\nicefrac{3}{2}$ case, with the various diagnostics close to the $\alpha=\nicefrac{3}{2}$ counterpart. In contrast, there are significant differences between the $\sigma'(z) \geq 0$ simulations and the $\alpha=\nicefrac{1}{2}$ simulations.
	 In the  $\sigma'(z) \geq 0$ series, the ratio of energy in the zonal mode to total energy continues to increase as $\keps/k_r$ is increased, whereas it asymptotes to a constant in the $\alpha=\nicefrac{1}{2}$ series. 
	Moreover, the ratio of domain average zonal speed to domain averaged meridional speed, $\overline{|u|}/\overline{|v|}$, is generally larger in the $\sigma \geq 0$ series than in the $\alpha=\nicefrac{1}{2}$ series.  Finally, for the largest values of $\keps/k_r$, the product $L_j\, \krh$ reaches smaller values in the $\sigma \geq 0$ simulations than in the $\alpha=\nicefrac{1}{2}$ simulations.

	\begin{figure}
  \centerline{\includegraphics[width=.8\columnwidth]{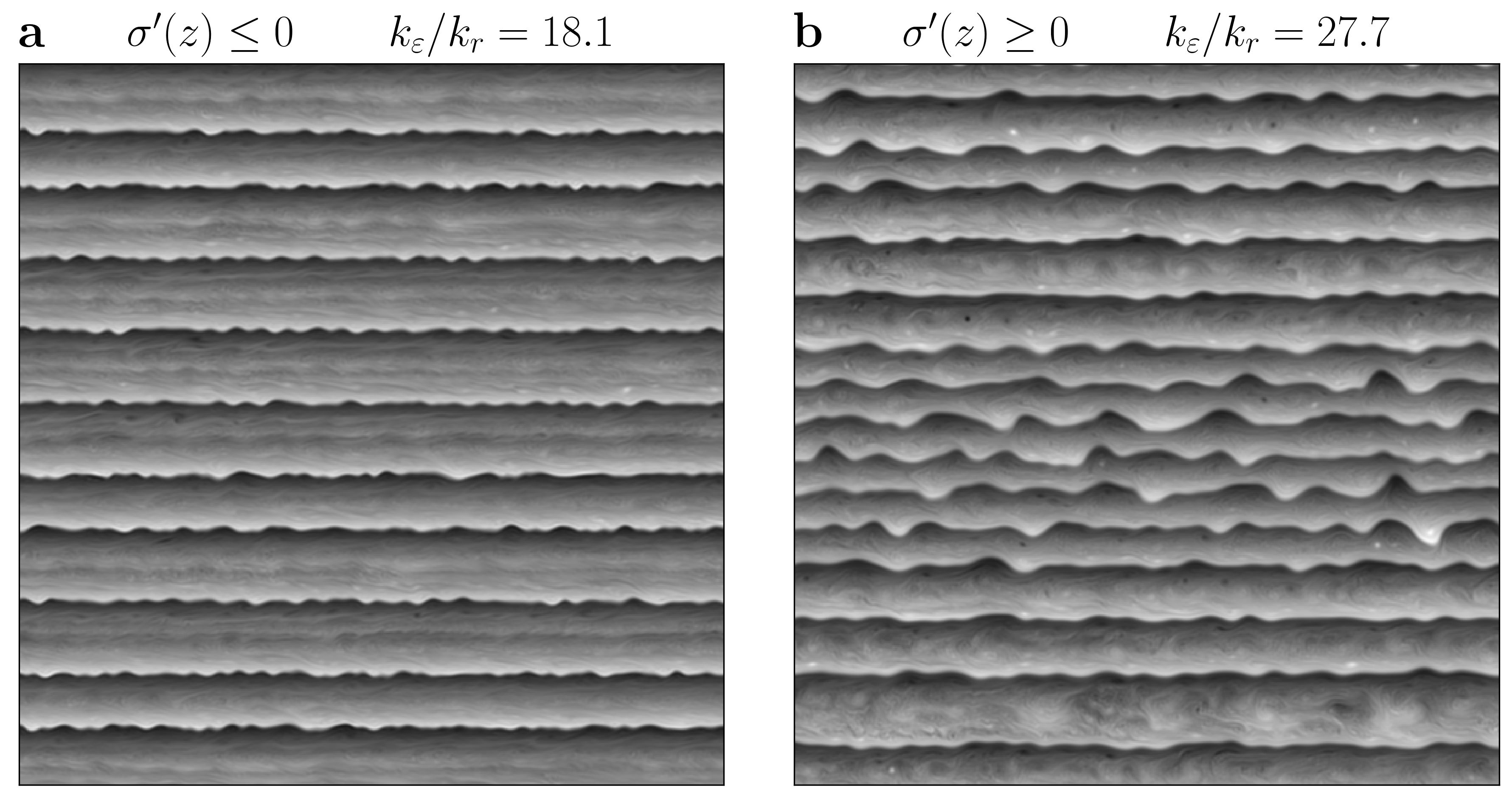}}
  \caption{Snapshots of the relative surface potential vorticity, $\theta$, normalized by its maximum value in the snapshot,  for simulations with inversion functions shown in figure \ref{F-mk_sigma_alt}.  Only one quarter of the domain is shown (i.e., $512^2$ grid points). }
  \label{F-alpha_jets_mixed}
   \end{figure}
   
   These differences can be explained by the snapshots of figure \ref{F-alpha_jets_mixed} as well as the zonal averages of figure \ref{F-model_staircase}. As expected from the model diagnostics, both the snapshots and the zonal average from the $\sigma^\prime \leq 0$  simulation are qualitatively similar to the $\alpha=\nicefrac{3}{2}$ simulation. In contrast, the $\sigma^\prime \geq 0$ snapshot is evidently closer to the staircase limit than the $\alpha=\nicefrac{1}{2}$ snapshot: the mixed zones are more homogeneous and the frontal zones are sharper. The zonal average of the $\sigma^\prime \geq 0$ simulation in figure \ref{F-model_staircase} also shows how the $\sigma^\prime \geq 0$ simulation is closer to the staircase limit than the $\alpha=\nicefrac{1}{2}$ simulation, although, again, zonal averaging in the presence of large amplitude undulations is artificially smoothing the jets. Therefore, the differences in the diagnostics between the $\sigma^\prime \geq 0$ series and the $\alpha=\nicefrac{1}{2}$ series stem from the more rapid approach (i.e., at smaller $\keps/k_r$) of the $\sigma^\prime \geq 0$ series to the staircase limit.

   \subsection{Simulations with fixed parameters} 
   
   The dependence of the non-dimensional number $\keps/k_r$ on the external parameters $\varepsilon, \Lambda$, and $r$ depends on the functional form of $m(k)$. For example, if $m(k) \sim k^\alpha$, then 
   \begin{equation}\label{eq:J_power}
   	 \keps/{k_r} = |\Lambda|^\frac{4-\alpha}{(2\alpha+1)(\alpha+2)}\, \varepsilon^\frac{\alpha-1}{(1+2\alpha)(\alpha+2)} \, r^\frac{-1}{\alpha+2}.
   \end{equation}
	Because the forcing intensity wavenumber, $\keps$, is obtained by solving the implicit equation for $\keps$ \eqref{eq:keps_condition}, an analogous expression for $\keps/k_r$ is not possible for general $m(k)$. However, at sufficiently large $\keps$, the inversion function asymptotes to $m(\keps) \approx \keps/\sigma_0$ and so, using $\alpha$-turbulence expression for $\keps$ \eqref{eq:keps_alpha} with $\alpha=1$, we obtain
	\begin{equation}\label{eq:J_strat}
		\keps/k_r \approx  |\Lambda|^\frac{\alpha}{\alpha+2} \, \varepsilon^\frac{1-\alpha}{3\alpha + 6} \, r^\frac{-1}{\alpha+2}\, m_0^\frac{1}{\alpha+2} \, \sigma_0^{2/3}
	\end{equation}
	for large $\keps$, where $\alpha$ is the approximate power law dependence of $m(k)$ near $k_r$. 
	%Note that the two expression have the opposite dependence on the energy injection rate $\varepsilon$; in $\alpha$-turbulence \eqref{eq:J_power} a larger $\varepsilon$ increases $\keps/k_r$ whereas for real stratification profiles \eqref{eq:J_strat} a larger $\varepsilon$ decreases $\keps/k_r$. For constant stratification ($\alpha=1$), $\keps/k_r$ is independent of $\varepsilon$ in both cases. 
	
	Therefore, simulations with identical $\keps/k_r$ but distinct inversion functions cannot be directly compared because they have different values of $\Lambda$ and $r$. 
	Here, we investigate how the stratification modifies jet structure as all other parameters are held fixed.
	We therefore run two additional series of simulations with the stratification profiles and inversion functions shown in figure \ref{F-mk_sigma}. 
	The stratification profiles were chosen so that they both have identical stratification at the upper boundary. One case corresponds to an increasing stratification profile, $\sigma^\prime \geq 0$, with an approximate power law dependence of $m(k) \sim k^{0.49}$ at small wavenumbers.  The second case consists of a decreasing stratification profile, $\sigma^\prime \leq 0$, with a $m(k) \sim k^{1.50}$ at small wavenumbers.
	Aside from the different stratification profiles, these two series of simulations are run under the same conditions as the constant stratification ($\alpha=1$) simulations of section \ref{SS-sims}, with identical values of $\Lambda$, $\varepsilon$, and $r$.
	
	\begin{figure}
  \centerline{\includegraphics[width=1\columnwidth]{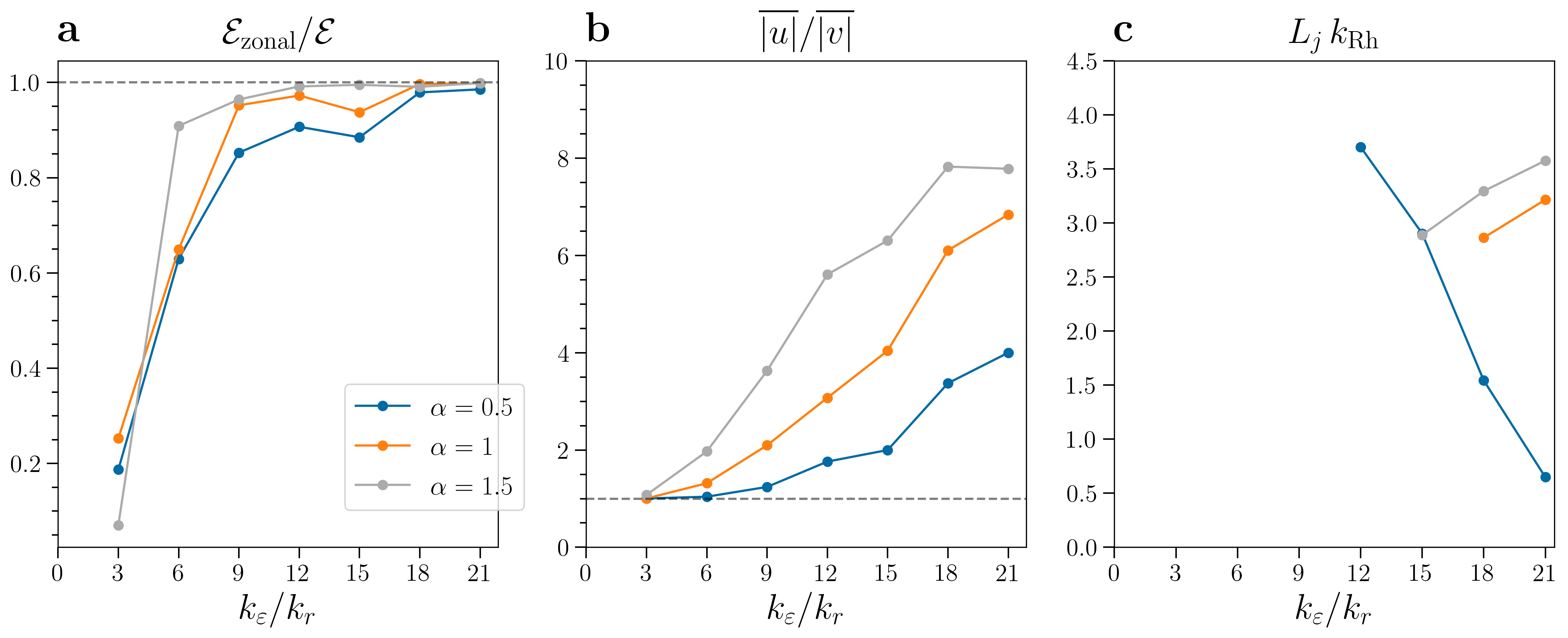}}
  \caption{As in figure \ref{F-krh_velocity}, but the $\sigma'\geq 0$ and $\sigma'\leq 0$ series now only differ from the $\sigma'=0$ (i.e., $\alpha=1$) series only in the vertical stratification (and hence the inversion function).}
  \label{F-krh_velocity_fixed}
   \end{figure}
	
\begin{figure}
  \centerline{\includegraphics[width=.8\columnwidth]{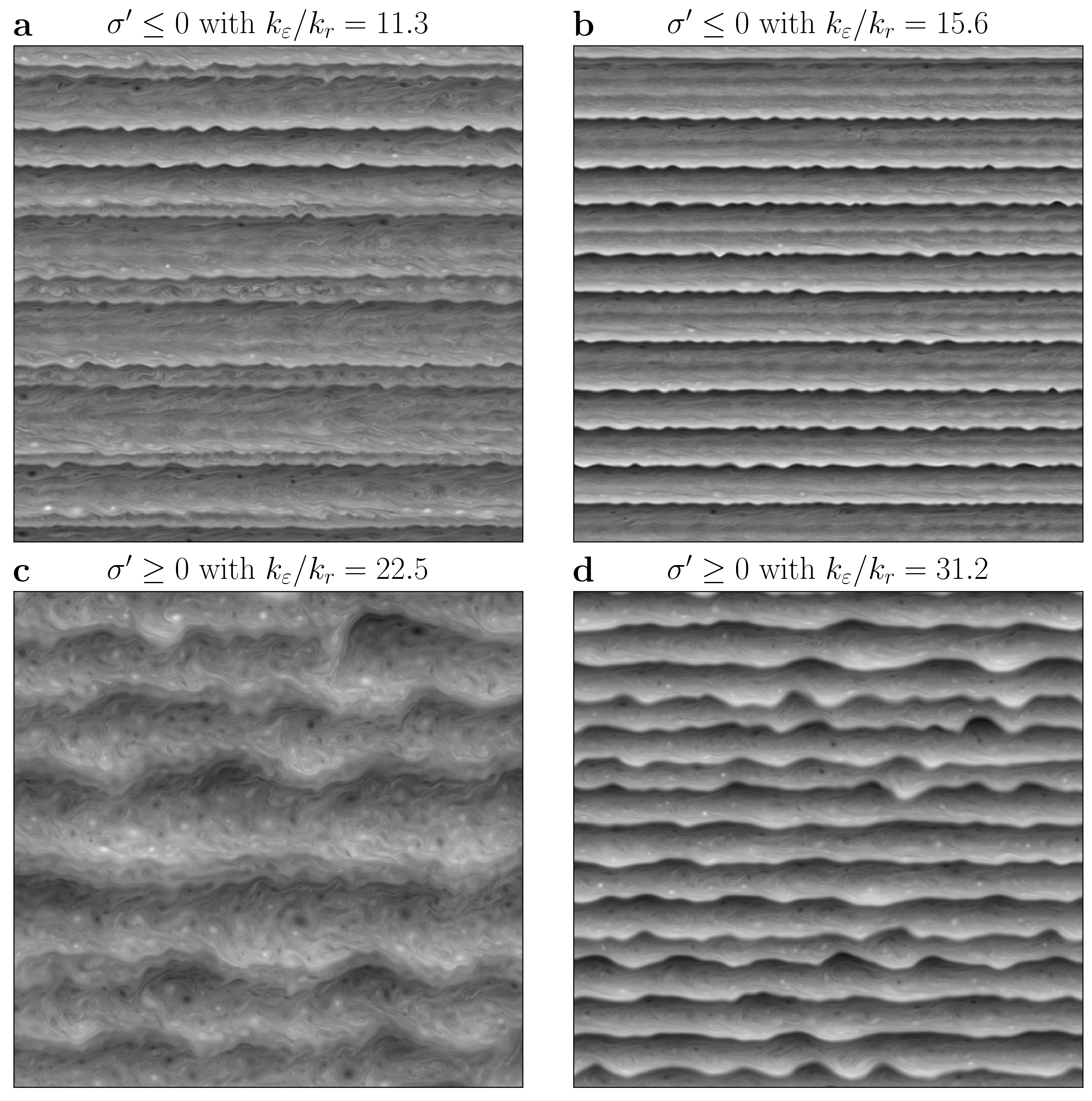}}
  \caption{Snapshots of relative surface potential vorticity, $\theta$, where $\theta$ is normalized by its maximum value in the snapshot. Panels (a) and (c) are from simulations with identical $\Lambda$, $r$, and $\varepsilon$ as the $\alpha=1$ simulation shown in figure \ref{F-alpha_jets}(c), whereas panels (b) and (d) are from simulations with identical $\Lambda$, $r$, and $\varepsilon$ as the $\alpha=1$ simulation shown in figure \ref{F-alpha_jets}(d). Only one quarter of the domain is shown (i.e., $512^2$ grid points).}
  \label{F-alpha_jets_fixed}
   \end{figure}

	Summary diagnostics are shown in figure \ref{F-krh_velocity_fixed}. We see that, at a fixed value of $\Lambda$ and $r$, more of the total energy is in the zonal mode in the $\sigma'(z) \leq 0$ simulation than in the constant stratification simulation, which in turn is larger than the $\sigma'(z) \geq 0$ simulation (and similarly for the ratio of area averaged zonal to meridional speeds, $\overline{\abs{u}}/\overline{\abs{v}}$). 
	Therefore, increased non-locality (larger  $\alpha$) promotes anisotropy in the velocity field and leads to larger zonal velocities relative to meridional velocities. Indeed, figure \ref{F-alpha_jets_fixed} shows $\theta$ snapshots from these simulations; the more local, $\sigma'\geq 0$, simulations have larger meridional undulations along the jets.
	Moreover, compared to the $\keps/k_r= 15$ constant stratification simulation in figure \ref{F-alpha_jets}(c), the $\sigma'(z) \leq 0$ simulation in figure \ref{F-alpha_jets_fixed}(a) is closer to the staircase limit whereas the frontal zones in the $\sigma'(z) \geq 0$ simulation [figure \ref{F-alpha_jets_fixed}(c)] remain broad. Finally, we show values of the product $L_j \, \krh$, relating the Rhines wavenumber to the half spacing between the jets,  in figure \ref{F-krh_velocity_fixed}(c). These values are similar to those in shown in figure \ref{F-krh_velocity}(c).
	 
\section{Conclusion} \label{S-conclusion}

We have examined the emergence of staircase-like buoyancy structures in surface quasigeostrophic turbulence with a mean background buoyancy gradient.
We found that the stratification's vertical structure controls the locality of the inversion operator and the dispersion of surface-trapped Rossby waves. 
As we go from decreasing stratification profiles [$\sigma'(z) \leq 0$] to increasing stratification profiles [$\sigma'(z) \geq 0$], the inversion operator becomes more local and Rossby wave less dispersive. 
In all cases, we find that the non-dimensional ratio, $\keps/k_r$, controls the extent of inhomogeneous buoyancy mixing. Larger $\keps/k_r$ correspond to sharper buoyancy gradients at jet centres with larger peak jet velocities that are separated by more homogeneous mixed-zones.
Moreover, we found that the staircase limit is reached at smaller $\keps/k_r$ in more non-local flows; the staircase limit is reached by $\keps/k_r \approx 15$ for our $\sigma \leq 0$ simulations compared to $\keps/k_r \approx 25$ for our $\sigma \geq 0$ simulations.

In addition, once the staircase limit is reached, the dynamics of the jets depends on the locality of the inversion operator and, hence, on the stratification's vertical structure. In flows with a more non-local inversion operator [or decreasing stratification, $\sigma'(z) \leq 0$], we obtain straight jets that are perturbed by dispersive, eastward propagating, along jet waves. In contrast, for more local flows [or over increasing stratification, $\sigma'(z) \geq 0$], we obtain jets with latitudinal meanders on the order of the jet spacing. The shape of these jets evolves in time as these meanders propagate westwards as weakly dispersive waves. 

The inversion operator's locality is also reflected in two more aspects of the dynamics. First, the domain-averaged zonal speed exceeds the domain-averaged meridional speed by approximately a factor of eight in our most non-local simulations, whereas this ratio is merely two in our most local simulations. This observation is consistent with the fact that jets are narrower and exhibit larger latitudinal meanders in more local flows. Second, for a given Rhines wavenumber, jets in more local flows are closer together. Indeed, we found $L_j \, \krh \approx 3-4$ in our most non-local simulations, where $L_j$ is the jet half spacing, as compared to $L_j \, \krh \approx 0.5-1.5$ in our most local simulations.

Several open questions remain. First, we have not examined the dynamics of the along jet waves. As we observed, these waves propagate eastwards in our most non-local simulations [with $\sigma'(z) \leq 0$] but westwards for our most local simulations [with $\sigma'(z) \geq 0$]. These waves are not described by the dispersion relation \eqref{eq:dispersion}; rather, the relevant model is that of freely propagating edge waves along a buoyancy discontinuity \citep{mcintyre_potential-vorticity_2008}. However, the difficulty here is that a jump discontinuity in the buoyancy field results in infinite velocities over constant or increasing stratification. In addition, the relationship of the along jet waves in the staircase limit to the non-linear zonons found by \cite{sukoriansky_nonlinear_2008} remains unclear.

The divergence of the velocity at a buoyancy discontinuities raises a second question. Is there a limit to how close the staircase limit can be approached?
In barotropic dynamics, the velocity remains finite at a jump continuity in the vorticity, and, in this case, \cite{scott_structure_2012} report that a vorticity staircase case can be approached arbitrarily. Whether this result continues to hold for arbitrarily sharp buoyancy gradients and arbitrarily large zonal velocities is not clear. Because the rms velocity seems to converge for arbitrarily sharp staircases, even for the most local inversion relations we considered, there may not be any energetic reason precluding arbitrarily sharp buoyancy gradients.

Finally, there remains the question of how relevant these results are for the upper ocean, which, in addition to surface buoyancy gradients, has interior potential vorticity gradients as well. In particular, our neglect of the $\beta$-effect limits the direct relevance of this model to the upper ocean. Whether surface buoyancy staircases can emerge under more realistic oceanic conditions requires further investigation. 

%\backsection[Supplementary data]{\label{SupMat}Supplementary material and movies are available at \\https://doi.org/10.1017/jfm.2019...}

\backsection[Acknowledgements]{The author thanks Stephen Griffies for useful discussions throughout the project as well as helpful comments that improved the clarity of manuscript.}

\backsection[Funding]{This report was prepared by Houssam Yassin under award NA18OAR4320123 from the National Oceanic and Atmospheric Administration, U.S. Department of Commerce. The statements, findings, conclusions, and recommendations are those of the author and do not necessarily reflect the views of the National Oceanic and Atmospheric Administration, or the U.S. Department of Commerce.}

\backsection[Declaration of interests]{ The author reports no conflict of interest.}

\backsection[Data availability statement]{The data that support the findings of this study are available within the article.}

\backsection[Author ORCID]{H. Yassin, https://orcid.org/0000-0003-1742-745X}

\bibliographystyle{jfm}
\bibliography{references}
%Use of the above commands will create a bibliography using the .bib file. Shown below is a bibliography built from individual items.

%\bibliographystyle{jfm}
%\bibliography{jfm2esam}

%\bibliography{references}

%% End of file `jfm2esam.bib'.

\end{document}